\documentclass[11pt]{article}
\usepackage{mydef2col}
\usepackage{markArticle}

\title{A GLE--LSV model with state-dependent leverage \\ and colored noise}

\shorttitle{A GLE--LSV model}

\author{
    \authorstyle{Andrey Itkin}
    \newline \newline
    \institution{FRE department, Tandon School of Engineering, New York University, email: \url{aitkin@nyu.edu}} \\
}

\date{\today}

\begin{document}

\maketitle

\lettrineabstract{We study a local stochastic volatility (LSV) model in which the variance factor follows a generalized Langevin equation (GLE) with colored noise, and past returns act on variance through a memory kernel with state-dependent amplitudes. For kernels given by finite sums of exponentials, we prove existence, uniqueness, and non-explosion under a growth condition linking the volatility map to the potential. In a one-mode family with a polynomial potential, feedback through log returns can cause explosion exactly when the squared volatility factor grows faster than the restoring force. Under structural and nondegeneracy conditions, an exact reduced Markovian state has a smooth density solving a deterministic forward equation. The Dupire conditional expectation therefore requires no stochastic partial differential equation. Under a bounded Sharpe ratio, exponential utility selects the minimal entropy martingale measure, whose premium solves a quadratic backward stochastic differential equation. Under a regularity hypothesis, which we verify in special cases with and without leverage, the premium depends only on the current state, and the squared VIX equals the expected average future local stochastic variance. Finally, a regularized calibration problem with bid--ask constraints admits a minimizer whenever a feasible point has finite objective value.}

%
%

\section{Introduction} \label{intro}

Local stochastic volatility (LSV) models combine a stochastic variance factor with a local function of time and price, chosen so that the model reproduces the observed surface of vanilla option prices, \cite{LiptonVolSmile2002, RenMadanQian2007}. The local function follows from Markovian projection. By the mimicking theorems of \cite{Gyongy1986, BrunickShreve2013}, the model has the market marginals when the squared local function, multiplied by the conditional expectation of the stochastic variance given the spot, equals the Dupire local variance, \cite{Dupire1994}. This conditional expectation depends on the law of the solution, so exact calibration is a nonlinear fixed-point problem of McKean--Vlasov type. Particle methods solve it numerically, \cite{GuyonHenryLabordere2011}, existence is known only in special cases, \cite{LackerShkolnikovZhang2020}, and optimal transport formulations replace the fixed point by an optimization problem, \cite{GuoLoeperWang2022}.

In most LSV models the stochastic factor is a Markov diffusion with a single time scale. Empirical volatility, by contrast, has memory over a wide range of horizons. Its persistence is the most visible feature. Past returns also act on future volatility asymmetrically, which is the leverage effect, and this effect decays on a time scale of its own, \cite{BouchaudMataczPotters2001}. In addition, the relation between returns and volatility is not symmetric under time reversal, \cite{Zumbach2009}. There is no reason for the persistence of volatility and the leverage effect to share the same time scales.

Rough volatility models describe these features through a fractional kernel with a single exponent, \cite{GatheralJaissonRosenbaum2018}. This exponent governs the kernel both at short lags, where it sets the roughness of volatility, and at long lags, where it sets the decay of the memory, and estimates of short-time roughness are sensitive to sampling and estimation, \cite{ContDas2024}. Fractional models are not Markovian. They could be approximated, for instance, by using the Dobric--Ojeda process, \cite{DobricOjeda2009, CarrItkin2019, Itkin2024FMF}, or via the fractional kernel represented by finite sums of exponentials that leads to Markovian systems with ordinary forward equations, \cite{AbiJaberElEuch2019, BayerBreneis}. Without such an approximation, the calibration condition of a rough LSV model involves the conditional law of the spot given the variance path, which solves a stochastic partial differential equation (SPDE) or rough partial differential equation (RPDE), \cite{BankBayerFrizPelizzari2025, BrigoLucic2026}.

The generalized Langevin equation (GLE) offers a different route. It describes a variable driven by a restoring force, a friction with memory, and a random force with nonzero correlation time, known as colored noise, \cite{Pavliotis2014}. When the memory kernel is a finite sum of exponentials, the GLE has an exact Markovian representation in which auxiliary variables carry the memory, \cite{OttobrePavliotis2011}. Applied to volatility, the GLE allows separate exponents for the short-time and long-time behavior of the kernel, and so decouples the scaling of volatility from its memory, \cite{ItkinGLE1}. Memory with state-dependent strength appears in the Marketron models of price formation, \cite{HalperinItkinMarketron, ItkinLeanMarketron}, and state-dependent friction with matched multiplicative noise has been studied in physics, \cite{LimWehrLewenstein2020}.

There are also economic and empirical reasons to let the strength of the leverage effect depend on the state. The balance-sheet leverage of financial intermediaries is procyclical, and margins and haircuts tighten in stress, which amplifies shocks through fire sales and liquidity spirals, \cite{Geanakoplos2010, AdrianShin2010, BrunnermeierPedersen2009}. Balance-sheet leverage is not the leverage effect of volatility modeling, which is the response of volatility to past returns, and the present model has no balance-sheet state. Its procyclicality suggests, however, that this response is stronger in stress. Empirically, the correlation between price and variance shocks becomes more negative at higher variance levels, \cite{BandiReno2012}. This motivates leverage amplitudes that depend on the variance state, without identifying them with that instantaneous correlation. Constant leverage already produces amplification and a skewed return distribution, so state dependence refines the model by letting the strength of the response change with the level of volatility.

Pricing in these models takes place in an incomplete market, because the variance driver is not traded. Exponential utility provides a selection rule that leads to the minimal entropy martingale measure, \cite{Frittelli2000, DelbaenEtAl2002, Becherer2003, HendersonHobson2009}. As applied to GLE, this approach is developed in \cite{HalperinItkinMarketron2,ItkinLeanMarketron}. In a Brownian filtration this measure is described by a quadratic backward stochastic differential equation (BSDE), \cite{Kobylanski2000, HuImkellerMuller2005, ManiaSchweizer2005}, and in Markovian stochastic volatility models by a semilinear partial differential equation, \cite{BenthKarlsen2005}. A single pricing measure for index and volatility derivatives is needed for the joint calibration to SPX and VIX options, which existing models address with rough or path-dependent volatility, \cite{GatheralJusselinRosenbaum2020, GuyonLekeufack2023}.

This paper studies an LSV model in which the variance driver follows a GLE with colored noise. We call it the GLE--LSV model. Past log returns act on the variance through a second memory kernel, the leverage kernel, whose decay rates are fixed and whose amplitudes depend on the current variance state. The model therefore separates the intrinsic persistence of volatility from the influence of past returns, and it lets the strength of that influence change with the level of volatility. A companion paper \cite{ItkinKazbekGLE} studies a GLE volatility model with constant leverage amplitudes and without a local function, and develops its estimation and joint historical SPX and SPX--VIX derivatives calibration. The colored-noise specification, the allocation of the return--variance correlation across memory modes, and the inventory-based premium used below were developed in that joint work.

Leverage memory built from log returns creates a feedback of its own. The It\^o drift of the log-price contains minus one half of the local stochastic variance. The leverage kernel passes this drift to the variance driver, where it acts against the restoring force and can make the variance explode. The local function changes the strength of this feedback but is not needed for it. A leverage kernel acting on arithmetic returns avoids this drift and needs a weaker condition for existence, but our proof of the true martingale property of the index needs the same growth condition for both conventions. We therefore keep log returns.

Our first results concern this feedback. For memory kernels given by finite sums of exponentials, we prove existence, uniqueness, non-explosion and the strong Markov property under a growth condition that links the volatility map to the GLE potential. We show that some growth control is needed. In a one-mode family with a polynomial potential, the growth of the restoring force is the exact boundary. The solution is global for every initial state when the squared volatility factor grows no faster, and it explodes with positive probability from suitable initial states when it grows faster. For potentials of degree higher than two, the general growth condition is therefore sufficient but not necessary.

Our second group of results concerns the structure of the state. When the physical drift depends on the friction and noise variables of each memory mode only through their difference, these variables combine, and the state dimension drops by one per mode. Under nondegeneracy conditions the reduced state has a smooth density that solves a deterministic forward equation. The conditional expectation in the calibration condition is then a ratio of integrals of this density, and no SPDE is needed. We also show where state dependence can enter the memory without destroying this finite representation or the covariance of the colored noise.

Our third group of results concerns pricing and calibration. Under a bounded Sharpe ratio, exponential utility selects the minimal entropy martingale measure, and its premium solves a quadratic BSDE. When the associated semilinear equation has a bounded classical solution with bounded exposures, the premium depends only on the current state, the discounted index is a true martingale, and the squared VIX equals the expected average future local stochastic variance. We prove this regularity in two cases, without leverage when the unhedgeable noise has a single direction or the memory modes are uncorrelated with returns, and with constant leverage in a family where a linear combination of the state variables is an autonomous Ornstein--Uhlenbeck factor. Otherwise we state it as a hypothesis. Finally, we pose the calibration of the local function as a regularized problem with SPX bid--ask constraints and VIX pricing errors, and prove that it has a minimizer whenever a feasible point has finite objective value. A short Monte Carlo experiment illustrates the leverage mechanism.

All results are exact for memory kernels given by finite sums of exponentials. Convergence to models with singular kernels, statistical estimation and calibration to market data are not treated here.

The paper is organized as follows. \Cref{sec:model} specifies the structural kernels and the physical dynamics. \Cref{sec:lift} states the assumptions and proves well-posedness, the necessity of a growth condition, the exact state reduction, and smoothness of the law. \Cref{sec:statedep} analyzes finite closure and the covariance restriction for state-dependent memory. \Cref{sec:pricing} constructs admissible pricing measures and identifies the utility-selected measure. \Cref{sec:localvol} treats Markovian projection, the VIX and its sensitivities, and existence for the regularized inverse problem. \Cref{sec:brigolucic} relates the density formulation to rough LSV, \cref{sec:numexp} illustrates the leverage mechanism numerically, and \cref{sec:conclusion} discusses the scope of the results and further questions.

\section{The GLE--LSV model} \label{sec:model}

Fix a horizon $T^\ast > 0$ and a filtered probability space $(\Omega,\mathcal F,(\mathcal F_t)_{t\in[0,T^\ast]},\PP)$ carrying independent standard Brownian motions $B_0, B_1, \ldots, B_N$ and an $\mathcal F_0$-measurable random initial state independent of them. The filtration is the augmentation of the one generated by the initial state and the Brownian motions. Let $S_t$ denote the index and $X_t = \log S_t$ its log-price. The variance is represented through a scalar driver $Y_t\in\mathbb R$ and a strictly increasing map $\phi:\mathbb R \to (0,\infty)$, with variance factor $V_t = \phi(Y_t)$ and volatility factor $f(V_t)$ for a positive function $f$, typically $f(v) = \sqrt v$. We write $F = f\circ\phi$. The short rate $r_t$ and the dividend yield $d_t$ are deterministic continuous functions. Expectations under the physical measure $\PP$ and a pricing measure $\QQ$ are denoted by $\EE^{\PP}$ and $\EE^{\QQ}$.

\subsection{Memory kernels} \label{sec:model_kernels}

\begin{definition}[Complete monotonicity] \label{def:cm}
A function $K:(0,\infty) \to \mathbb R$ is completely monotone if it is infinitely
differentiable and $(-1)^n K^{(n)}(t)\ge 0$ for every integer $n\ge 0$ and every $t > 0$.
\end{definition}

By the Bernstein--Widder theorem, $K$ is completely monotone if and only if there is a unique positive Borel measure $\mu_K$ on $[0,\infty)$ with
\begin{equation} \label{eq:bernstein}
K(t) = \int_{[0,\infty)} e^{-\gamma t}\,\mu_K(d\gamma), \qquad t > 0,
\end{equation}
\cite{Widder1941, SchillingSongVondracek2012}. The variable $\gamma$ is a relaxation rate and $\mu_K$ is the spectral measure. Following \cite{ItkinGLE1}, the intrinsic memory has separate short-time and long-time exponents $\beta_0, \beta_\infty \in (0,1)$, and a crossover rate $\kappa > 0$ marks the transition between them.

\begin{definition}[Structural kernel] \label{def:structural}
The structural kernel is
\begin{equation} \label{eq:kernel_product}
\Kyy(t) = c\,t^{-\beta_0}(1  +  \kappa t)^{\beta_0 - \beta_\infty}, \qquad c > 0.
\end{equation}
\end{definition}
For either ordering of the exponents, $\Kyy(t)\sim c\,t^{-\beta_0}$ as $t \to  0$ and $\Kyy(t)\sim c\kappa^{\beta_0 - \beta_\infty}t^{-\beta_\infty}$ as $t \to \infty$. A pure power, and hence the fractional kernel of rough volatility, is the case $\beta_0 = \beta_\infty$.

\begin{lemma}[Spectral density] \label{lem:densities}
The structural kernel is completely monotone, and $\mu_K(d\gamma) = q_K(\gamma)\,d\gamma$ with
\begin{equation} \label{eq:structural_density}
q_K(\gamma) = \frac{c\,\kappa^{\beta_0 - \beta_\infty}}{\Gamma(\beta_\infty)}\,\gamma^{\beta_\infty - 1}e^{-\gamma/\kappa}\,{}_1F_1\Big(\beta_0;\beta_\infty;\frac{\gamma}{\kappa}\Big) > 0, \qquad \gamma > 0,
\end{equation}
where ${}_1F_1$ is Kummer's confluent hypergeometric function. As $\gamma \to \infty$, $q_K(\gamma)\sim c\,\gamma^{\beta_0 - 1}/\Gamma(\beta_0)$, and as $\gamma \to  0$, $q_K(\gamma)\sim c\kappa^{\beta_0 - \beta_\infty}\gamma^{\beta_\infty - 1}/\Gamma(\beta_\infty)$.
\end{lemma}

\begin{proof}
The series ${}_1F_1(a;b;x) = \sum_{n\ge 0}(a)_nx^n/((b)_nn!)$ has positive terms for $a,b,x > 0$, so $q_K > 0$. For $b > 0$ and $z > 1$,
\[ \int_0^\infty e^{-zs}s^{b - 1}{}_1F_1(a;b;s)\,ds = \Gamma(b)z^{-b}(1 - z^{-1})^{-a},
\]
\cite{OlverEtAl2010}. The substitution $\gamma = \kappa s$ with $z = 1  +  \kappa t$, $a = \beta_0$ and $b = \beta_\infty$ gives
\[ \int_0^\infty e^{-\gamma t}q_K(\gamma)\,d\gamma = c\,t^{-\beta_0}(1  +  \kappa t)^{\beta_0 - \beta_\infty} \]
for every $t > 0$, which is \cref{eq:bernstein}. The asymptotics follow from ${}_1F_1(a;b;x) \to  1$ as $x \to  0$ and ${}_1F_1(a;b;x)\sim\Gamma(b)e^xx^{a - b}/\Gamma(a)$ as $x \to \infty$.
\end{proof}

\begin{myremark} \label{rem:kernel_additive}
For $\beta_0 > \beta_\infty$, the additive kernel $c_0t^{-\beta_0}e^{-\kappa t}  +  c_\infty t^{-\beta_\infty}$ with $c_0,c_\infty > 0$ has the same two regimes. It is completely monotone, with spectral density
\[
q(\gamma) = \frac{c_\infty\gamma^{\beta_\infty - 1}}{\Gamma(\beta_\infty)}  +  \mathbf 1_{\{\gamma > \kappa\}}\frac{c_0(\gamma - \kappa)^{\beta_0 - 1}}{\Gamma(\beta_0)},
\]
since $t^{-\beta} = \Gamma(\beta)^{-1}\int_0^\infty\gamma^{\beta - 1}e^{-\gamma t}\,d\gamma$ and multiplication by $e^{-\kappa t}$ shifts the spectral variable by $\kappa$. The density is positive on $(0,\infty)$: the first term is positive for all $\gamma>0$, and the second is positive on $(\kappa,\infty)$, where it has the singularity $(\gamma-\kappa)^{\beta_0-1}$ at $\gamma=\kappa$, which is integrable because $\beta_0>0$. The product form \cref{eq:kernel_product} covers both orderings with one formula, and the results below apply to either kernel.
\end{myremark}

\begin{proposition}[Finite kernels and approximation] \label{prop:finite} \NL
\begin{enumerate}
\item[(i)] Let $a_1,\ldots,a_n > 0$ and let $\gamma_1,\ldots,\gamma_n\ge 0$ be distinct. Then $K(t) = \sum_ja_je^{-\gamma_jt}$ if and only if $\mu_K = \sum_ja_j\delta_{\gamma_j}$, and then its Laplace transform $\hat K(z) = \sum_ja_j/(z  +  \gamma_j)$ is rational with simple poles at $-\gamma_j$ and residues $a_j$.

\item[(ii)] A completely monotone kernel is a finite exponential sum if and only if its spectral measure is finitely atomic. In particular, the structural kernel of \cref{def:structural}, the additive kernel of \cref{rem:kernel_additive}, and the fractional kernel are not finite exponential sums.

\item[(iii)] Let $K$ be completely monotone and finite on $(0,\infty)$. For every compact interval $I\subset(0,\infty)$ and $\varepsilon > 0$ there is a positive exponential sum $K_n$ with $\sup_{t\in I}|K(t) - K_n(t)| < \varepsilon$.
\end{enumerate}
\end{proposition}

\begin{proof} \NL
(i) Inserting $\mu = \sum_ja_j\delta_{\gamma_j}$ in \cref{eq:bernstein} gives the sum, and the converse follows from uniqueness of $\mu_K$. Integration term by term gives $\hat K$.

(ii) The first claim is uniqueness of $\mu_K$. The spectral measures in \cref{lem:densities,rem:kernel_additive}, and the density $\gamma^{\beta - 1}/(\Gamma(\beta)\Gamma(1 - \beta))$ of the fractional kernel, are absolutely continuous and nonzero, hence not atomic.

(iii) Let $I = [t_0,t_1]$. For $R > 0$, $\int_{(R,\infty)}e^{-\gamma t}\mu_K(d\gamma) \le  e^{-Rt_0/2}K(t_0/2)$ for $t\ge t_0$, so truncation of $\mu_K$ to $[0,R]$ costs at most $\varepsilon/2$ on $I$ for large $R$. The truncated measure is finite, and it is the weak limit of positive atomic measures supported in $[0,R]$, for example Riemann sums. The family $\{\gamma\mapsto e^{-\gamma t}: t\in I\}$ is uniformly bounded and uniformly equicontinuous on $[0,R]$, so the weak convergence is uniform in $t\in I$.
\end{proof}

\begin{myremark} \label{rem:fractional}
For the fractional kernel $K_\beta(t) = t^{-\beta}/\Gamma(1 - \beta)$, $\beta\in(0,1)$, the Laplace transform is $\hat K_\beta(z) = z^{\beta - 1}$, analytic on $\mathbb C\setminus(-\infty,0]$ with a branch cut along the negative axis. Stieltjes inversion, $q_\beta(\gamma) = -\pi^{-1}\lim_{\varepsilon\downarrow 0}\operatorname{Im}\hat K_\beta(-\gamma  +  i\varepsilon)$, returns the spectral density $q_\beta(\gamma) = (\sin\pi\beta/\pi)\,\gamma^{\beta - 1}$. A rational transform has only poles, so an $n$-term sum replaces the cut by $n$ points and can follow the kernel only for lags whose dominant rates $\gamma\sim 1/t$ lie in the sampled band of rates. Since $K_\beta(\lambda t) = \lambda^{-\beta}K_\beta(t)$, every decade of lags carries the same structure, which is why the number of terms grows with $\log(T/t_0)$.
\end{myremark}

The approximation in (iii) holds only on a compact lag interval. A finite sum decays like its slowest exponential at large lags and is bounded at zero, whereas the structural kernels have an algebraic tail and a singular origin. For a target accuracy $\varepsilon$ on $[t_0,T]$, near-optimal exponential sums use $N = O(\log(1 / \varepsilon)( \log(1/\varepsilon)  +  \log(T/t_0)))$ terms, \cite{BeylkinMonzon2005,BeylkinMonzon2010}.

\begin{definition}[Approximating kernel] \label{def:approx}
The intrinsic kernel used in the dynamics is
\begin{equation} \label{eq:kyy}
\Kyy^{(N)}(\tau) = \sum_{i=1}^N a_ie^{-\gamma_i\tau}, \qquad a_i > 0, \qquad \gamma_i > 0,
\end{equation}
obtained by positive quadrature of $q_K$ on a truncated rate interval. The exponents, the crossover, and the amplitudes are the structural parameters. The distinct pairs $(a_i,\gamma_i)$ are numerical quadrature parameters and are not fitted independently.
\end{definition}

\begin{table}[!htb]
\centering
\caption{Comparison of the fractional kernel of rough volatility and the structural kernel \cref{eq:kernel_product}.}
\label{tab:kernels}
\begin{tabular}{lcc}
 \toprule
 & Fractional (rough) kernel & Structural kernel \\
\midrule
Form & $t^{-\beta}/\Gamma(1-\beta)$ & $c\,t^{-\beta_0}(1 + \kappa t)^{\beta_0-\beta_\infty}$ \\[2pt]
Short-time exponent & $\beta$ & $\beta_0$ \\[2pt]
Long-time exponent & $\beta$ & $\beta_\infty$ \\[2pt]
Spectral density $q(\gamma)$ & $\dfrac{\sin\pi\beta}{\pi}\,\gamma^{\beta-1}$ & $\dfrac{c\,\kappa^{\beta_0-\beta_\infty}}{\Gamma(\beta_\infty)}\,\gamma^{\beta_\infty-1}e^{-\gamma/\kappa}\,{}_1F_1\!\Big(\beta_0;\beta_\infty;\tfrac{\gamma}{\kappa}\Big)$ \\[6pt]
Scale invariance & $K(\lambda t)=\lambda^{-\beta}K(t)$ & none unless $\beta_0=\beta_\infty$ \\[2pt]
Singular at $t=0$ & yes & yes \\[2pt]
Algebraic tail at $t=\infty$ & yes & yes \\[2pt]
Finite exponential sum & no & no \\[2pt]
Free kernel parameters & $1$ ($\beta$) & $3$ ($\beta_0,\beta_\infty,\kappa$) plus $c$ \\
\bottomrule
\end{tabular}
\end{table}

\Cref{tab:kernels} collects the differences between the fractional kernel of rough volatility and the structural kernel \cref{eq:kernel_product}. The fractional kernel is the special case $\beta_0=\beta_\infty$ of the structural kernel, in which the crossover factor $(1 + \kappa t)^{\beta_0-\beta_\infty}$ collapses to $1$. Both kernels are completely monotone, both have a singular origin, and both have an algebraic tail. The structural kernel differs in one essential respect: it decouples the short-time exponent $\beta_0$ from the long-time exponent $\beta_\infty$ through the crossover rate $\kappa$. A single Hurst index, as in the fractional specification, imposes a common exponent on the short-time singularity of the kernel, which sets the roughness of volatility, and on its long-time decay. The structural kernel relaxes this constraint at the cost of two additional shape parameters: an independent long-time exponent and the crossover rate $\kappa$. The overall scale $c$ is counted separately. Its spectral measure, given in \cref{lem:densities}, is a two-power-law density with a crossover at $\gamma\approx\kappa$, whereas the fractional spectral measure is a single power. Neither kernel is a finite exponential sum, by \cref{prop:finite}(ii).

The exact finite-dimensional theory developed below therefore applies to the quadrature approximation $\Kyy^{(N)}$ of \cref{def:approx}, not to the kernels themselves. Convergence of the finite approximation to either kernel as $N \to \infty$ is a separate question, which is not addressed here.

\subsection{Dynamics under the physical measure} \label{sec:model_P}

Under $\PP$ the index is a LSV diffusion,
\begin{equation} \label{eq:price_P}
\frac{dS_t}{S_t} = \mu_t\,dt  +  L(t,S_t)F(Y_t)\,dW^X_t, \qquad W^X = \sum_{i=1}^N \rho_i B_i  +  \sqrt{1 - \rho^2}\,B_0,
\end{equation}
where $L$ is the local volatility function, $\mu_t$ is the physical drift, and the modal correlations satisfy $\sum_i\rho_i^2 = \rho^2 < 1$, so that $W^X$ is a Brownian motion with $d\langle W^X,B_i\rangle_t = \rho_i\,dt$. A convenient allocation is
\begin{equation} \label{eq:rho_alloc}
\rho_i = \rho\sqrt{p_i}, \qquad p_i = \frac{\gamma_ia_i/(\gamma_i^2  +  \lambda_c^2)}{\sum_{k=1}^N\gamma_ka_k/(\gamma_k^2  +  \lambda_c^2)}, \qquad \lambda_c > 0,
\end{equation}
but the results below hold for any allocation with $\sum_i\rho_i^2 = \rho^2$.

The variance driver follows the GLE
\begin{equation} \label{eq:gleY_lev}
\dot Y_t  +  \int_0^t\Kyy^{(N)}(t - s)\dot Y_s\,ds  +  \sum_{i=1}^N e^{-\gamma_it}u_{i,0} = -U'(Y_t) - \sum_{j=1}^Mc_j(Y_t)g_{j,t}  +  \xi_t,
\end{equation}
with unit instantaneous friction, a confining potential $U$\footnote{The function $U$ plays the role of a potential for the restoring force $-U'(Y_t)$ in the drift of the driver. No physical interpretation of $U$ as an energy is required.}, and leverage states
\begin{equation} \label{eq:gstates}
g_{j,t} = e^{-\lambda_jt}g_{j,0}  +  \int_0^te^{-\lambda_j(t - s)}\,dX_s, \qquad \lambda_j > 0.
\end{equation}

Each $g_j$ is an exponentially weighted sum of past log returns with decay rate $\lambda_j$. The constants $u_{i,0}$ and $g_{j,0}$ carry the memory accumulated before time zero. The leverage term equals $-\int_0^t\Kyx(t - s;Y_t)\,dX_s$, up to its initial part, for the leverage kernel
\vspace{-0.5\baselineskip}
\begin{equation} \label{eq:kyx}
\Kyx(\tau;y) = \sum_{j=1}^Mc_j(y)e^{-\lambda_j\tau}.
\end{equation}
Its rates are fixed and its amplitudes depend on the current state of the driver. The amplitude multiplies an already defined leverage state, so the term is not an anticipative integral. With $c_j\ge 0$, a negative return raises the drift of $Y$. The leverage states use log returns. Leverage states built from arithmetic returns, $dS_t/S_t = dX_t  +  \tfrac12L(t,S_t)^2F(Y_t)^2\,dt$, are a common alternative, \cite{GuyonLekeufack2023}, and the two conventions have different well-posedness theories. The force is $\xi_t = \sum_i\zeta_{i,t}$, where
\begin{equation} \label{eq:ou_modes}
d\zeta_{i,t} = -\gamma_i\zeta_{i,t}\,dt  +  q_i\,dB_{i,t}, \qquad q_i = \sqrt{2\Theta a_i\gamma_i}, \qquad \Theta > 0.
\end{equation}

The noise modes share the rates of the friction modes, and the loadings $q_i$ are chosen so that the covariance of $\xi$ is $\Theta\Kyy^{(N)}$ under stationary initialization. The relation covers the memory part of the friction only. The instantaneous friction has no matching white force, so the stationary law of the driver, when it exists, need not be proportional to $e^{-U/\Theta}$. The price coordinate carries no memory of its own in this baseline. Its diffusion coefficient is fixed by the LSV structure, not by a covariance restriction.

\begin{myremark} \label{rem:twod}
In the symmetric formulation of \cite{ItkinGLE1}, both the log-price and the driver obey a GLE with a matrix kernel $(K_{ab})_{a,b\in\{X,Y\}}$, and each row of the kernel is tied to an additive stationary Gaussian force. Applied to the price, such a relation would prescribe the diffusion of the traded index through a memory kernel. In an LSV model that diffusion is $L(t,S)F(Y)$, multiplicative and state dependent, and it is constrained by absence of arbitrage and by the option surface. The present model therefore keeps the GLE structure, with its noise relation, for the non-traded driver only. The price-side entries $K_{XX}$ and $K_{XY}$ do not enter as friction paired with a random force, although memory in the physical drift of returns is not excluded. The return-to-variance entry $\Kyx$ is retained as a directed coupling. The direction of this coupling is consistent with the time-reversal asymmetry known as the Zumbach effect.
\end{myremark}

\section{Markovian representation and well-posedness} \label{sec:lift}

\subsection{The lifted state} \label{sec:model_lift}

For the intrinsic memory, define friction states
\begin{equation} \label{eq:intrinsic_lift}
u_{i,t} = e^{-\gamma_it}u_{i,0}  +  a_i\int_0^te^{-\gamma_i(t - s)}\,dY_s, \qquad i = 1,\ldots,N,
\end{equation}
so that
\[ \sum_iu_{i,t} = \sum_ie^{-\gamma_it}u_{i,0}  +  \int_0^t\Kyy^{(N)}(t - s)\dot Y_s\,ds
\]
is the memory friction in \cref{eq:gleY_lev}. The lifted state is
\begin{equation} \label{eq:liftstate}
\Zs_t = (X_t, Y_t, u_{1,t}, \ldots,u_{N,t}, \zeta_{1,t}, \ldots,\zeta_{N,t}, g_{1,t}, \ldots,g_{M,t}) \in \mathbb R^{2  +  2N  +  M},
\end{equation}
with generic point $\mathbf z = (x,y,u,\zeta,g)$. Write
\begin{equation} \label{eq:driver_drift}
m(\mathbf z) = -U'(y) - \sum_{i=1}^Nu_i  +  \sum_{i=1}^N\zeta_i - \sum_{j=1}^M c_j(y) g_j, \qquad \varsigma(t,\mathbf z) = \Lh(t,x)F(y),
\end{equation}
for the driver drift and the effective index volatility, with $\Lh(t,x) = L(t,e^x)$. The lifted system under $\PP$ is
\begin{alignat}{2} \label{eq:lift_P}
dX_t &= \big(\mu_t - \tfrac12\varsigma_t^2\big)\,dt  +  \varsigma_t\,dW^X_t, &\qquad dY_t &= m(\Zs_t)\,dt, \\
du_{i,t} &= \big(a_im(\Zs_t) - \gamma_iu_{i,t}\big)\,dt, &\qquad d\zeta_{i,t} &= -\gamma_i\zeta_{i,t}\,dt  +  q_i\,dB_{i,t}, \nonumber \\
dg_{j,t} &= -\lambda_jg_{j,t}\,dt  +  dX_t, \nonumber
\end{alignat}
for $i = 1,\ldots,N$ and $j = 1,\ldots,M$, where $\varsigma_t = \varsigma(t,\Zs_t)$.

\begin{theorem}[Markovian representation] \label{thm:lift}
Let $X$ be a continuous semimartingale satisfying the first equation in \cref{eq:lift_P}, let $Y$ be continuous and adapted with continuously differentiable paths, and define $u$, $\zeta$, and $g$ by \cref{eq:intrinsic_lift,eq:ou_modes,eq:gstates}. Then $(X,Y)$ satisfies \cref{eq:gleY_lev} if and only if $\Zs$ solves \cref{eq:lift_P}.
\end{theorem}

\begin{proof}
Since $e^{\gamma_it}u_{i,t} = u_{i,0}  +  a_i\int_0^te^{\gamma_is}\,dY_s$, the product rule gives $du_{i,t} = a_i\,dY_t - \gamma_iu_{i,t}\,dt$. The same computation with the deterministic factor $e^{\lambda_jt}$, for which no covariation term arises, gives $dg_{j,t} = -\lambda_jg_{j,t}\,dt  +  dX_t$. Substituting the expression for $\sum_iu_{i,t}$ into \cref{eq:gleY_lev} shows that the equation is equivalent to $\dot Y_t = m(\Zs_t)$, and then $du_{i,t} = (a_im(\Zs_t) - \gamma_iu_{i,t})\,dt$. Conversely, the linear equations for $u_i$ and $g_j$ with given initial values have the unique solutions \cref{eq:intrinsic_lift,eq:gstates}, by variation of constants.
\end{proof}

The driver $Y$ has no direct Brownian forcing. Its paths are continuously differentiable at fixed $N$, so any power-law behavior is confined to the resolved band of lags and is not a statement about the regularity of paths.

\subsection{Standing assumptions} \label{sec:assumptions}

Write $U_\ast = \inf_y U(y)$, $\Lh(t,x) = L(t,e^x)$, and define on the state space
\begin{equation} \label{eq:lyapunov}
G(\mathbf z) = 1  +  U(y) - U_\ast  +  \sum_{i=1}^N\frac{u_i^2}{2a_i}  +  \sum_{i=1}^N \zeta_i^2  +  \sum_{j=1}^Mg_j^2  +  x^2.
\end{equation}

\begin{assumption} \label{ass:std} \NL
\begin{enumerate}
\item[(A1)] $U\in C^2(\mathbb R)$, and there are $k_1 > 0$, $k_2\ge 0$ with $-yU'(y) \le  -k_1y^2  +  k_2$ for all $y$.

\item[(A2)] $\phi$ is continuous and strictly increasing with $\inf\phi > 0$, and $F = f\circ\phi$ is locally Lipschitz with $F\ge f_{\min} > 0$ and
\begin{equation} \label{eq:growth_link}
F(y)^4 \le  C_F\big(1  +  U(y) - U_\ast\big), \qquad y\in\mathbb R.
\end{equation}

\item[(A3)] Each $c_j:\mathbb R \to [0,c_{\max}]$ is locally Lipschitz.

\item[(A4)] $L$ is Borel, $\ell_{\min} \le  L \le \ell_{\max}$ for constants $0 < \ell_{\min} \le \ell_{\max}$, and $|\Lh(t,x) - \Lh(t,x')| \le  K_L|x - x'|$ for all $t,x,x'$.

\item[(A5)] The physical drift is $\mu_t = \mu(t,\Zs_t)$ for a Borel function $\mu$ on $[0,T^\ast]\times\mathbb R^{2  +  2N  +  M}$, locally Lipschitz in the state uniformly in $t$, with a bounded market price of traded risk: $|\mu(t,\mathbf z) - (r_t - d_t)| \le \bar\theta\,\varsigma(t,\mathbf z)$ for a constant $\bar\theta$, where $\varsigma = \Lh F$ is defined in \cref{eq:driver_drift}.

\item[(A6)] The initial state $\Zs_0$ is $\mathcal F_0$-measurable and satisfies $\EE^{\PP}[G(\Zs_0)] < \infty$ for the function $G$ of \cref{eq:lyapunov}.
\end{enumerate}
\end{assumption}

By (A1), $U(y)\ge U_\ast  +  \tfrac{k_1}{4}y^2 - C$ for a constant $C$, so $U$ is coercive. Condition \cref{eq:growth_link} links the volatility map to the potential. It holds for $f(v) = \sqrt v$, $\phi(y) = v_{\min}  +  \tfrac12(y  +  \sqrt{y^2  +  \epsilon^2})$ with $v_{\min},\epsilon > 0$, and every potential satisfying (A1), since then $F^4 = \phi^2 \le  C(1  +  y^2)$. A quartic potential admits faster growth, $F(y)^4 \le  C(1  +  y^4)$. For $f(v)=\sqrt v$, the lognormal map $\phi(y) = e^y$ fails the uniform lower bound in (A2), and with a polynomial potential it also violates \cref{eq:growth_link} as $y \to  + \infty$. The latter restriction depends on the potential and does not exclude every exponentially growing volatility map.

\subsection{Existence, uniqueness, and the Markov property} \label{sec:wellposed}

\begin{theorem}[Well-posedness] \label{thm:wellposed}
Under \cref{ass:std}, the system \cref{eq:lift_P} has a unique strong solution on $[0,T^\ast]$, and it does not explode. There is a constant $C_G$, depending only on the constants in \cref{ass:std}, on $N$, $M$, on $q_1,\ldots,q_N$, and on $\sup_{t \le  T^\ast}|r_t - d_t|$, such that
\begin{equation} \label{eq:moment_bound}
\EE^{\PP}[G(\Zs_t)] \le \EE^{\PP}[G(\Zs_0)]\,e^{C_Gt}, \qquad t\in[0,T^\ast].
\end{equation}
The index $S_t = e^{X_t}$ and the variance factor $V_t = \phi(Y_t)\ge\inf\phi$ are strictly positive, and $\Zs$ is a time-inhomogeneous strong Markov process.
\end{theorem}

\begin{proof}
See \cref{app:theorem2}.
\end{proof}

\begin{myremark} \label{rem:growth}
The bound in \cref{eq:growth_link} controls feedback through log returns. The drift of $X$ contains $-\tfrac12\varsigma^2$, so the leverage contribution to the driver includes $\tfrac12\sum_jc_j(Y_t)\int_0^te^{-\lambda_j(t-s)}\varsigma_s^2\,ds$. This term has the opposite sign to a restoring force, and the bound on $2g_j(\mu - \tfrac12\varsigma^2)$ in the proof requires control of $\varsigma^4$.
\end{myremark}

\begin{proposition}[Noise covariance] \label{prop:fdt}
Let the $\zeta_{i,0}$ be independent, independent of the Brownian motions, with $\zeta_{i,0}\sim\mathcal N(0,\Theta a_i)$. Then, for $s,t\ge 0$,
\begin{equation} \label{eq:noise_covariance}
\EE^{\PP}[\zeta_{i,t}\zeta_{j,s}] = \delta_{ij}\Theta a_ie^{-\gamma_i|t - s|}, \qquad \EE^{\PP}[\xi_t\xi_s] = \Theta\Kyy^{(N)}(|t - s|).
\end{equation}
The law of $(\zeta_1,\ldots,\zeta_N)$ under $\PP$ does not depend on $U$, $L$, $\mu$, or the leverage amplitudes. The identity is a stationary one: it holds for all $s,t\ge0$ only under the stated initialization. For a non-stationary initialization of the $\zeta_{i,0}$, the covariance acquires a transient term that decays at rate $\gamma_i$.
\end{proposition}

\begin{proof}
The modes solve autonomous linear equations driven by independent Brownian motions, so they are independent and their law depends only on $(a_i,\gamma_i,\Theta)$. For $s \le  t$ we have
\[ \zeta_{i,t} = e^{-\gamma_i(t - s)}\zeta_{i,s}  +  q_i\int_s^te^{-\gamma_i(t - r)}\,dB_{i,r},
\qquad \operatorname{Var}(\zeta_{i,s}) = e^{-2\gamma_is}\Theta a_i  +  q_i^2(1 - e^{-2\gamma_is})/(2\gamma_i) = \Theta a_i,
\]
since $q_i^2 = 2\Theta a_i\gamma_i$ by \cref{eq:ou_modes}. Hence $\EE^{\PP}[\zeta_{i,t}\zeta_{i,s}] = \Theta a_ie^{-\gamma_i(t - s)}$. Summing over modes gives the second identity.
\end{proof}

Independent noises are essential here. A common Brownian motion with the same loadings would produce cross-covariances $q_iq_j/(\gamma_i  +  \gamma_j)$ at equal times and would not reproduce \cref{eq:noise_covariance}. The identity describes the force alone. The friction states depend on the past of $Y$ and do not have this covariance. In particular, the friction states $u_i$ cannot be given an independent Gaussian initialization analogous to that of the noise modes. A joint stationary initialization of the intrinsic subsystem, when it exists, must respect their dependence on the driver and its past.

\subsection{The growth condition and the return convention} \label{sec:convention}

Leverage states built from arithmetic returns replace $dX_t$ by $dS_t/S_t = dX_t  +  \tfrac12\varsigma_t^2\,dt$ in \cref{eq:gstates}, so that in \cref{eq:lift_P} they follow $dg_{j,t} = (\mu_t - \lambda_jg_{j,t})\,dt  +  \varsigma_t\,dW^X_t$.

\begin{proposition}[Arithmetic returns] \label{prop:arithmetic}
For arithmetic returns, assume \cref{ass:std} with \cref{eq:growth_link} replaced by $F(y)^2 \le C_F(1 + U(y) - U_\ast)$. Then the system has a unique strong solution on $[0,T^\ast]$, which does not explode and is strong Markov, and \cref{eq:moment_bound} holds for the function obtained from $G$ by replacing $x^2$ with $\sqrt{1 + x^2}$.
\end{proposition}

\begin{proof}
Replace $x^2$ in \cref{eq:lyapunov} by $\sqrt{1  +  x^2}$, whose first and second derivatives are bounded by one. The $x$-terms of $\mathcal L^{\PP}G$ are then bounded by $C(1  +  \varsigma^2)$. The leverage terms become $2g_j(\mu - \lambda_jg_j)  +  \varsigma^2$, and $2g_j\mu \le  g_j^2  +  C(1  +  \varsigma^2)$ by (A5). The weaker growth condition gives $\varsigma^2 \le  CG$, so $\mathcal L^{\PP}G \le  C_GG$. Since $G\ge c|\mathbf z| - C$, the rest of the proof of \cref{thm:wellposed} applies.
\end{proof}

The next theorem shows that some growth condition is necessary for log returns, identifies the sharp growth exponent in a one-mode family, and compares the two conventions. The potential has degree $m\ge 2$ on the positive axis, which covers the cubic--quartic potentials of \cite{ItkinGLE1} with $m = 4$.

\begin{theorem}[Necessity of a growth condition] \label{thm:dichotomy}
Take $N=M=1$, $a_1=a>0$, $\gamma_1=\lambda_1=\lambda>0$, $\rho_1=0$, $\Theta>0$, $c_1(y)=c\ge 0$, $L\equiv1$, $\mu=0$, and zero short rate and dividend yield. Let $U$ satisfy (A1) and, for constants $m\ge 2$, $k>0$, $K>0$ and all $y\ge 1$,
\[
ky^m \le  U(y) - U_\ast  +  K, \qquad U'(y) \le  Ky^{m-1}, \qquad U''(y) \le  Ky^{m-2}.
\]
Let $f(v)=\sqrt v$ and $\phi=F^2$, where $F\in C^1(\mathbb R)$ is strictly increasing with $\inf F>0$, and let $C_1y^p \le  F(y)^2 \le  C_2y^p$ for $y\ge 1$, with constants $0<C_1 \le  C_2$ and $p>0$.
\begin{enumerate}
\item[(i)] If $p \le  m/2$, the solution is global for every initial state satisfying (A6). The same holds if $p \le  m-1$ and, for some $\bar y\ge 1$, $U''(y)\ge ky^{m-2}$ for $y\ge\bar y$ and $U''(y)\ge -K$ for $y \le \bar y$.

\item[(ii)] If $c>0$ and $p>m-1$, then for every $T\in(0,T^\ast]$ and every $R>0$ there are deterministic initial states from which the maximal solution explodes before $T$ with probability at least $1-e^{-R/(2c)}$.

\item[(iii)] With arithmetic returns, the solution is global for every initial state satisfying (A6) if $p \le  m$.

\item[(iv)] If $c=0$, the solution is global for either convention.
\end{enumerate}
\end{theorem}

\begin{proof}
See \cref{app:theorem3}.
\end{proof}

For $m = 2$, parts (i) and (ii) meet at $p = 1$, so \cref{eq:growth_link} is sharp in the exponent. For $m>2$, the lower bound on $U''$ in (i) holds for every polynomial potential of degree $m$ that satisfies (A1), and (i) and (ii) then meet at $p=m-1$. With $c>0$, log-return feedback is therefore global for every initial state if and only if $p \le  m-1$. Condition \cref{eq:growth_link} corresponds to $p \le  m/2$, so in this family it is sufficient but not necessary when $m>2$. For a cubic--quartic potential, $m = 4$. Log-return feedback then explodes when $F^2$ grows faster than $y^3$ and is global when $F^2$ grows at most like $y^3$, and arithmetic returns are global up to growth like $y^4$.

The proof of (ii) uses neither the upper bound on $F^2$ nor $\inf F > 0$. Part (ii) therefore holds for every $F\in C^1(\mathbb R)$ with $F^2$ nondecreasing and bounded below by a power above $m - 1$. The theorem uses constant leverage, a subclass of the allowed state-dependent amplitudes, to isolate the effect of the return convention. Stationary initialization of the intrinsic noise is also possible in (ii). Take $\zeta_0\sim\mathcal N(0,\Theta a)$ independent of the Brownian motions and $u_0=\zeta_0-h_0$ with $h_0$ fixed as in the proof. The same argument applies, and the covariance identity of \cref{prop:fdt} holds.

For arithmetic returns and $m>2$, the theorem does not cover the exponents $p>m$. The function $H$ in the proof of (i) relies on the closed pair $(Y,w)$, which exists because the friction and leverage rates coincide and the amplitude is constant. With several modes or distinct rates there is no such pair, and with a state-dependent amplitude the drift of $w$ acquires the term $-c'(Y)\,g\,\dot Y$. Whether \cref{eq:growth_link} can be weakened in \cref{thm:wellposed} in these cases remains open.

\subsection{Reduction of the Markovian state} \label{sec:reduced}

\begin{proposition}[Reduced lift] \label{prop:reduced}
Assume that $\mu$ depends on $(u,\zeta)$ only through $h = \zeta - u$ (equivalently, $\mu(t,\mathbf z) = \tilde\mu(t,x,y,h,g)$ for a function $\tilde\mu$). Then $\Zs$ solves \cref{eq:lift_P} if and only if $\Zs^{\mathrm r} = (X,Y,h_1,\ldots,h_N,g_1,\ldots,g_M)$, with $h_{i,t} = \zeta_{i,t} - u_{i,t}$, solves
\begin{alignat}{2} \label{eq:reduced_lift}
dX_t &= \big(\mu_t - \tfrac12\varsigma_t^2\big)\,dt  +  \varsigma_t\,dW^X_t, &\qquad dY_t &= \tilde m(\Zs^{\mathrm r}_t)\,dt, \\
dh_{i,t} &= -\big(\gamma_ih_{i,t}  +  a_i\tilde m(\Zs^{\mathrm r}_t)\big)\,dt  +  q_i\,dB_{i,t}, &\qquad dg_{j,t} &= -\lambda_jg_{j,t}\,dt  +  dX_t, \nonumber \\
\tilde m(x,y,h,g) &= -U'(y)  +  \sum_ih_i - \sum_jc_j(y)g_j, \nonumber
\end{alignat}
and $u$ is recovered from $u_0$ and $Y$ by \cref{eq:intrinsic_lift}. Under \cref{ass:std}, $\Zs^{\mathrm r}$ is a strong Markov process with values in $\mathbb R^{2  +  N  +  M}$.
\end{proposition}

\begin{proof}
The drift $m$ depends on $(u,\zeta)$ only through $\sum_i(\zeta_i - u_i) = \sum_ih_i$, so $m(\Zs) = \tilde m(\Zs^{\mathrm r})$. Subtracting the equations for $u_i$ and $\zeta_i$, which share the rate $\gamma_i$, gives the equation for $h_i$. Conversely, given a solution of \cref{eq:reduced_lift}, define $u$ by \cref{eq:intrinsic_lift} and $\zeta = h  +  u$. Then $d\zeta_{i,t} = - \gamma_i \zeta_{i,t}\,dt  +  q_i\,dB_{i,t}$ and $\Zs$ solves \cref{eq:lift_P}. Uniqueness and the Markov property transfer through this pathwise bijection and \cref{thm:wellposed}.
\end{proof}

This is the quasi-Markovian form of a GLE with an exponential kernel, \cite{OttobrePavliotis2011, Pavliotis2014}. The reduction uses the common decay rates of friction and noise, but not the amplitude restriction $q_i^2 = 2\Theta a_i\gamma_i$. The combined state $h_i$ includes friction feedback and does not have the covariance \cref{eq:noise_covariance}. The separate coordinates are kept in the statements that concern the noise, and the reduced state is the one used for computation. No minimality claim is made among all possible representations. For example, a leverage mode with $c_j\equiv0$ can be omitted when the physical drift does not depend on that mode and the quantities of interest do not require its history.

The reduction is conditional on the drift depending on the friction and noise coordinates only through $h$. If the physical drift depends on $u$ and $\zeta$ separately, the state dimension reverts to $2 + 2N + M$. Under a pricing measure, the same restriction must hold for the pricing drift, including the premium, for the reduction to survive the change of measure.

\subsection{Smoothness of the law and the forward equation} \label{sec:hypo}

The conditional expectation that calibrates the local function is a ratio of integrals of the joint density of the state. The driver carries no Brownian forcing, and the log-price and the leverage states share a single Brownian direction, so the generator is degenerate. Smoothness of the law follows from a bracket condition instead.

\begin{theorem}[Hypoellipticity] \label{prop:hypo}
Under \cref{ass:std}, assume that $\mu$ depends on $(u,\zeta)$ only through $h$, as in \cref{prop:reduced}. Let $|\rho| < 1$, let the rates $\lambda_1,\ldots,\lambda_M$ be distinct, and assume that $U'$, the $c_j$, $F$, $\mu$, and $\Lh$ are $C^\infty$ in $(t,\mathbf z)$. Then the generator $\mathcal L^{\mathrm r}$ of \cref{eq:reduced_lift} satisfies the parabolic H\"ormander condition at every point. Consequently, for $t > 0$ the law of $\Zs^{\mathrm r}_t$ has a density $p_t$, the map $(t,\mathbf z)\mapsto p_t(\mathbf z)$ is $C^\infty$ on $(0,T^\ast)\times\mathbb R^{2  +  N  +  M}$, and $p$ solves the forward equation $\partial_tp = (\mathcal L^{\mathrm r})^\ast p$ classically.
\end{theorem}

\begin{proof}
See \cref{app:theorem4}.
\end{proof}

The theorem replaces the assumption of a smooth joint density by a condition on the coefficients. Distinct leverage rates matter. If $\lambda_j=\lambda_k=\lambda$, then $g_j(t)-g_k(t)=e^{-\lambda t}(g_j(0)-g_k(0))$. Consequently the transition law from a fixed initial state is supported on a lower-dimensional set. For a random initial state, randomness in the initial differences can supply additional dimensions, so this conclusion need not hold for the unconditional law. For leverage states built from arithmetic returns, the drifts of $X$ and $g_j$ differ by the additional common term $\tfrac12\varsigma^2$. One further bracket then supplies the missing direction, which requires $F' > 0$. Smoothness does not give positivity of the spot marginal at every strike, which is needed for the denominator of the conditional expectation in the wings.

The bracket condition holds locally uniformly in the leverage rates. Consider compact sets of states and parameters on which the coefficients and their derivatives are bounded, the diffusion loadings and the rates are bounded below, and the rates have pairwise separation bounded below by a positive constant. There the determinant $\prod_{j<k}(\lambda_k-\lambda_j)$ is bounded below, and together with the bounds on the other entries it bounds the smallest singular value of the matrix $(p_l(\lambda_j))_{l,j}$ from below. If the rates coalesce, the determinant vanishes, the span of the brackets degenerates, and the local subelliptic estimates lose their uniformity. This local uniformity does not give global subelliptic estimates. In the limit of equal rates, the leverage states $g_j$ become linearly dependent through their initial differences, and the transition law may concentrate on a lower-dimensional set for a fixed initial state.

\subsection{Linear response} \label{sec:linear}

The memory affects the scaling of the driver on a band of frequencies. For the linearized intrinsic dynamics this can be made explicit. Write
\[
\widehat K(\omega) = \int_0^\infty\Kyy^{(N)}(t)e^{-i\omega t}\,dt = \sum_i a_i /(\gamma_i  +  i\omega).
\]

\begin{proposition}[Spectral density] \label{prop:spectrum}
Let $U(y) = \tfrac12a_Uy^2$ with $a_U > 0$ and $c_j\equiv 0$. Then the intrinsic subsystem $(Y,h)$ of \cref{eq:reduced_lift} has a unique stationary law, which is Gaussian, and in stationarity $Y$ has the spectral density
\begin{equation} \label{eq:friction_spectrum}
S_Y(\omega) = \frac{S_\xi(\omega)}{|a_U  +  i\omega(1  +  \widehat K(\omega))|^2}, \qquad S_\xi(\omega) = 2\Theta\operatorname{Re}\widehat K(\omega) = \sum_{i=1}^N\frac{2\Theta a_i\gamma_i}{\gamma_i^2  +  \omega^2}.
\end{equation}
\end{proposition}

\begin{proof}
The system for $(Y,u,\zeta)$ is linear and block triangular, with $\zeta$ driving $(Y,u)$. The $\zeta$-block has eigenvalues $-\gamma_i$. For the $(Y,u)$-block with $\zeta = 0$, put $A = a_UY  +  \sum_iu_i$. The function $E = \tfrac12a_UY^2  +  \sum_iu_i^2/(2a_i)$ satisfies $\dot E = -A^2 - \sum_i(\gamma_i/a_i)u_i^2$, which vanishes only at the origin, so by LaSalle's principle the block is asymptotically stable. The full drift matrix is therefore stable, and the linear system has a unique stationary Gaussian law. In stationarity, $Y$ is the output of the causal filter with transfer function $(a_U  +  i\omega(1  +  \widehat K(\omega)))^{-1}$ applied to $\xi$, as follows by Fourier transformation of \cref{eq:gleY_lev}. The modes $\zeta_i$ are independent stationary Ornstein--Uhlenbeck processes with spectral densities $2\Theta a_i\gamma_i/(\gamma_i^2  +  \omega^2) = 2 \Theta \operatorname{Re}(a_i /(\gamma_i  +  i\omega))$, and spectral densities multiply by the squared modulus of the transfer function.
\end{proof}

Suppose the quadrature resolves a structural kernel with $\widehat K(\omega)\approx c\,\Gamma(1 - \beta_0)(i\omega)^{\beta_0 - 1}$ on a band where $|\omega\widehat K(\omega)|$ dominates both $|\omega|$ and $a_U$. On that band \cref{eq:friction_spectrum} gives $S_Y(\omega)\asymp|\omega|^{-1 - \beta_0}$, an effective Hurst index $H = \beta_0/2$. A white force of the same scale would give $S_Y\asymp|\omega|^{-2\beta_0}$ and $H = \beta_0 - \tfrac12$. At high frequencies the unit instantaneous friction dominates, and at fixed $N$ the driver is continuously differentiable. A roughness estimate, a friction exponent, and a skew exponent are therefore not interchangeable without specifying the band of frequencies and the coupling to returns.

\section{Where state dependence can enter the memory} \label{sec:statedep}

The kernels of \cref{sec:model} carry memory of a fixed shape in the lag variable. This section asks where, and in what form, state dependence can be introduced without destroying the finite Markovian representation of \cref{thm:lift} or the noise covariance of \cref{prop:fdt}. In the GLE--LSV model the amplitudes of the leverage kernel depend on the driver through
\begin{equation} \label{eq:cj_form}
c_j(y) = \bar c_j\big(1  +  \eta\,\psi(y)\big),
\end{equation}
with $\bar c_j\ge 0$, a sensitivity $\eta\ge 0$, and a bounded locally Lipschitz function $\psi$ with $1  +  \eta\psi > 0$, for example $\psi(y) = \tanh((y - y_\star)/s_\star)$. At $\eta = 0$ the amplitudes are constant. A positive $\eta$ strengthens the whole leverage term structure in stress and weakens it in calm, while the fixed rates keep its shape in the lag variable. The amplitudes \cref{eq:cj_form} satisfy (A3).

\subsection{Forms that close in finite dimensions} \label{sec:statedep_ampl}

Current-state amplitudes need no additional states: the term $\sum_j c_j( Y_t)g_{j,t}$ is a function of $Y_t$ and the states \cref{eq:gstates}. Two further forms also close.

\begin{proposition}[Finite closure] \label{prop:closure}
Let $c_j$ satisfy (A3).
\begin{enumerate}
\item[(i)] \emph{Past-state amplitudes.} For a constant rate $\lambda_j > 0$, the process
\[ \tilde g_{j,t} = \int_0^te^{-\lambda_j(t - s)} c_j(Y_s)\,dX_s \quad \mathrm{solves} \quad d\tilde g_{j,t} = -\lambda_j\tilde g_{j,t}\,dt + c_j(Y_t)\,dX_t.
\]

\item[(ii)] \emph{Path-integrated rates.} Let $\lambda_j:\mathbb R \to [\lambda_{\min},\lambda_{\max}]$ be locally Lipschitz with $\lambda_{\min} > 0$, and put $E_{j,t} = \exp\big(\int_0^t\lambda_j(Y_r)\,dr\big)$. The process
\[
\hat g_{j,t} = E_{j,t}^{-1}\int_0^t E_{j,s}\,dX_s \quad \mathrm{solves} \quad d\hat g_{j,t} = -\lambda_j(Y_t) \hat g_{j,t}\,dt + dX_t.
\]
We write $\hat g_{j,t} = \int_0^t \exp\big(-\int_s^t \lambda_j(Y_r)\,dr\big)\,dX_s$ for this process, with the factor $E_{j,t}^{-1}$ taken outside the stochastic integral.
\end{enumerate}

If the leverage term in \cref{eq:gleY_lev} is replaced by $-\sum_j\tilde g_j$ or by $-\sum_j\bar c_j\hat g_j$ with constants $\bar c_j\ge 0$, the conclusions of \cref{thm:wellposed} remain valid with the same state dimension.
\end{proposition}

\begin{proof}
Apply the product rule to $e^{\lambda_jt}\tilde g_{j,t}$ and to $\hat g_{j,t} = E_{j,t}^{-1}\int_0^t E_{j,s}\,dX_s$, where $E_{j}$ is a continuous process of finite variation, so no covariation terms arise. For the last claim, the coefficients remain locally Lipschitz. In the proof of \cref{thm:wellposed}, the term $-2\lambda_j(y)g_j^2$ is nonpositive, and the replacement of $dX$ by $c_j(Y)\,dX$ multiplies the drift and quadratic variation inputs by at most $c_{\max}$ and $c_{\max}^2$. The bound $\mathcal L^{\PP}G \le  C_GG$ follows with a larger constant.
\end{proof}

State-dependent rates therefore do not by themselves destroy finite closure. The convention matters. The weight $e^{-\lambda_j(Y_t)(t - s)}$, which applies the current rate to the whole past, gives
\[
\int_0^t e^{-\lambda_j(Y_t)(t - s)}\,dX_s = \Upsilon_t( \lambda_j(Y_t)), \qquad  \Upsilon_t(\lambda) = \int_0^t e^{-\lambda(t - s)}\,dX_s.
\]
Its evaluation requires the whole family $\{ \Upsilon_t (\lambda): \lambda\in\lambda_j(\mathbb R) \}$, and it does not close with the states of (ii). Path-integrated rates close, but they lose the interpretation of a fixed function of the lag, so complete monotonicity and the exponents of a lag kernel cannot be invoked. Fixed rates with current-state amplitudes keep a common lag basis across regimes. This is the choice made here.

\subsection{The cost of state dependence in the intrinsic kernel} \label{sec:statedep_fdt}

Suppose instead that the intrinsic amplitudes depend on the state, $\Kyy(\tau;y) = \sum_ia_i(y)e^{-\gamma_i\tau}$. If one attempts to preserve the present identity by equating the unconditional force covariance directly to this random kernel, the proposed relation is
\begin{equation} \label{eq:state_fdt}
\EE^{\PP}[\xi_t\xi_s] = \Theta\sum_{i=1}^Na_i(Y_t)e^{-\gamma_i(t - s)}\quad\text{a.s.}, \qquad 0 \le  s \le  t.
\end{equation}

The unconditional covariance identity is restricted. Let $\gamma_1,\ldots,\gamma_N$ be distinct and let $\xi$ be square integrable with continuous covariance function. If \cref{eq:state_fdt} holds for some $t > 0$ and all $s\in[0,t]$, its left-hand side is a deterministic continuous function of $s$, while the functions $s\mapsto e^{-\gamma_i(t - s)}$ are linearly independent on $[0,t]$. Hence each coefficient $a_i(Y_t)$ is almost surely constant, and continuous amplitudes are constant on every interval in the support of the law of $Y_t$. This restricts \cref{eq:state_fdt} only. It is not a general impossibility statement for state-dependent intrinsic memory. State-dependent fluctuation--dissipation models with matched multiplicative noise and friction use a different relation, see \cite{LimWehrLewenstein2020}. Such formulations are outside the additive stationary-noise model studied here. If intrinsic friction amplitudes are changed while that noise is kept fixed, the covariance constraint and the parameter reduction described below are lost for the modified friction.

Observed persistence of volatility can arise from a slowly decaying memory or from a persistent random force, and different combinations of the two can describe the same data. By \cref{eq:noise_covariance}, the noise variances equal $\Theta a_i$, so a single scale $\Theta$ replaces $N$ free noise weights, with the same ratio of noise to friction on every memory mode. The relation therefore acts as a regularization of the physical-measure estimation and helps identify the kernel parameters. The price is a bias if the data do not satisfy the restriction. Neither stationarity nor the absence of arbitrage requires it.

\subsection{The leverage kernel carries no such constraint} \label{sec:statedep_which}

\begin{myremark}[The leverage entry] \label{rem:leverage_entry}
Under \cref{ass:std}, for any amplitudes satisfying (A3), the law of the force $\xi$ under $\PP$, and in particular \cref{eq:noise_covariance} under the stationary initialization of \cref{prop:fdt}, does not depend on the $c_j$, by the last statement of \cref{prop:fdt}. The martingale part of each $g_j$ is $\int_0^\cdot\varsigma_s\,dW^X_s$, which is spanned by the index. Consequently, under any equivalent change of measure, the leverage states have no independent Brownian component, and their diffusion is entirely along $W^X$. Exposures through the leverage states are therefore hedgeable using the index.
\end{myremark}

The leverage feedback is a directed forcing of the driver by past returns. It is not part of the friction--noise pair, so its amplitudes may depend on the state without affecting \cref{eq:noise_covariance}. The second fact in \cref{rem:leverage_entry} matters under the pricing measure, since exposures through the leverage states are hedgeable.

Note that the Marketron model of \cite{HalperinItkinMarketron, ItkinLeanMarketron} reduces to a GLE with a single memory mode whose amplitude depends on the state. In the present terms it is the case $M = 1$ of the mechanism in \cref{eq:cj_form}. The extension to several modes gives the leverage a term structure rather than a single relaxation time.

\section{The pricing measure} \label{sec:pricing}

\subsection{Equivalent martingale measures} \label{sec:model_Q}

The traded assets are the money market account and the index with dividend yield $d$. Let $\tilde S_t = \exp(-\int_0^t(r_s - d_s)\,ds)S_t$ be the discounted total-return index, so that
\[
d\tilde S_t/\tilde S_t = \bar\mu_t\,dt  +  \varsigma_t\,dW^X_t, \qquad \bar\mu_t = \mu_t - (r_t - d_t).
\]
The market price of traded risk is $\theta_t = \bar\mu_t/\varsigma_t$, and $|\theta| \le \bar\theta$ by (A5). Collect the Brownian motions in $\mathbf B = (B_1,\ldots,B_N,B_0)$ and let $\mathbf r = (\rho_1,\ldots,\rho_N,\sqrt{1 - \rho^2})$, a unit vector with $W^X = \mathbf r\cdot\mathbf B$. The variance driver is not traded, and the $N$ directions orthogonal to $\mathbf r$ are not spanned by the index.

\begin{proposition}[Martingale measures] \label{prop:emm} \NL
\begin{enumerate}
\item[(i)] Let $\boldsymbol\Lambda = (\Lambda_1,\ldots,\Lambda_N,\Lambda_0)$ be bounded and progressively measurable with $\mathbf r\cdot\boldsymbol\Lambda = 0$, and put $\boldsymbol\beta = -\theta\mathbf r  +  \boldsymbol\Lambda$. Then $\mathcal E(\int\boldsymbol\beta\cdot d\mathbf B)$ is a true martingale, and $\QQ^{\Lambda}$, defined by its terminal value, is equivalent to $\PP$. Under $\QQ^{\Lambda}$ the processes $B^{\QQ}_k = B_k - \int_0^\cdot\beta_k\,dt$ are independent Brownian motions, $\tilde S$ is a local martingale, and the lifted state solves
\begin{alignat}{2} \label{eq:lift_S}
dX_t &= \big(r_t - d_t - \tfrac12\varsigma_t^2\big)\,dt  +  \varsigma_t\,dB^{\QQ}_t, &\qquad B^{\QQ} &= \mathbf r\cdot\mathbf B^{\QQ}, \qquad dY_t = m(\Zs_t)\,dt, \\
du_{i,t} &= \big(a_im(\Zs_t) - \gamma_iu_{i,t}\big)\,dt, &\qquad d\zeta_{i,t} &= \big(-\gamma_i\zeta_{i,t} - q_i\rho_i\theta_t  +  q_i\Lambda_{i,t}\big)\,dt  +  q_i\,dB^{\QQ}_{i,t}, \nonumber \\
dg_{j,t} &= - \lambda_jg_{j,t}\,dt  +  dX_t. \nonumber
\end{alignat}

\item[(ii)] Conversely, let $\QQ\sim\PP$ coincide with $\PP$ on $\mathcal F_0$, and let $\tilde S$ be a local martingale under $\QQ$. Then the density process of $\QQ$ is $\mathcal E(\int\boldsymbol\beta\cdot d\mathbf B)$ for a progressively measurable $\boldsymbol\beta$ with $\int_0^{T^\ast}|\boldsymbol\beta_t|^2\,dt < \infty$ a.s. and $\mathbf r\cdot\boldsymbol\beta = -\theta$.
\end{enumerate}
\end{proposition}

\begin{proof} \NL
(i) Novikov's condition holds because $\boldsymbol\beta$ is bounded, and Girsanov's theorem gives the Brownian motions $B^{\QQ}_k$. Since
\[
\bm r \cdot \bm r = 1, \qquad \bm r \cdot \bm\Lambda = 0, \qquad dW^X = dB^{\QQ}  +  \bm r \cdot \bm\beta\,dt = dB^{\QQ} - \theta\,dt,
\]
so $d\tilde S/\tilde S = \varsigma\,dB^{\QQ}$. Substituting $dB_i = dB^{\QQ}_i  +  \beta_i\,dt$ in \cref{eq:lift_P} gives \cref{eq:lift_S}.

(ii) The density process is a positive martingale in a Brownian filtration with unit initial value. By the martingale representation theorem it equals $\mathcal E(\int\boldsymbol\beta\cdot d\mathbf B)$ for some $\boldsymbol\beta$ with $\int|\boldsymbol\beta|^2\,dt < \infty$. By Girsanov's theorem the $\QQ$-drift of $\tilde S/\tilde S$ is $\bar\mu  +  \varsigma\,\mathbf r\cdot\boldsymbol\beta$, which vanishes if and only if $\mathbf r\cdot\boldsymbol\beta = -\theta$.
\end{proof}

The traded index receives the risk-neutral drift, while the premium $\boldsymbol\Lambda$, the market price of memory risk, acts only on the noise modes. The finite-variation coordinates $Y$ and $u_i$ keep their pathwise expressions. The leverage states also keep their definition as exponentially weighted sums of past log returns, and they are semimartingales with the same Brownian term as $X$. The stationary covariance \cref{eq:noise_covariance} is a physical-measure property and is not preserved once state-dependent drifts are added to the noise modes.

The diffusion coefficient $\varsigma = LF$ of the index is the same under $\PP$ and under every $\QQ^{\Lambda}$, because an equivalent change of measure alters drifts only. The local function calibrated to option prices therefore also enters the physical dynamics, and the quadratic variation of $\log S$, observable under $\PP$, is $\int\varsigma^2\,dt$ under both measures. This is a restriction linking the two measures, and it is why the physical-measure estimation of the kernels and the potential is carried out for a given $L$.

\begin{lemma}[True martingale property] \label{lem:martingale}
Let $\boldsymbol\Lambda_t = \boldsymbol\Lambda(t,\Zs_t)$, where $\boldsymbol\Lambda(t,\cdot)$ is bounded by $\bar\Lambda$, locally Lipschitz uniformly in $t$, and orthogonal to $\mathbf r$. Under \cref{ass:std}, the system \cref{eq:lift_S} has a unique strong non-exploding solution under $\QQ^{\Lambda}$, which satisfies \cref{eq:moment_bound} with a constant depending also on $\bar\theta$ and $\bar\Lambda$, and $\tilde S/\tilde S_0$ is a true $\QQ^{\Lambda}$-martingale. If $\EE^{\QQ^{\Lambda}}[S_0] < \infty$, in particular for a deterministic initial state, $\tilde S$ is a true $\QQ^{\Lambda}$-martingale.
\end{lemma}

\begin{proof}
Relative to \cref{eq:lift_P}, the drifts of $X$ and $g_j$ in \cref{eq:lift_S} contain $r - d$ in place of $\mu$, and the noise modes acquire bounded drifts. Since $2\zeta_ib \le \zeta_i^2  +  b^2$ for bounded $b$, the proof of \cref{thm:wellposed} applies. For the martingale property, write $\mathcal E_t = \tilde S_t/\tilde S_0 = \mathcal E(\int\varsigma\,dB^{\QQ})_t$ and consider \cref{eq:lift_S} with $dB^{\QQ}_k$ replaced by $dB^S_k  +  \varsigma r_k\,dt$. This share-measure system has the additional drifts $\varsigma^2$ in $dX$ and $dg_j$ and $q_i\rho_i\varsigma$ in $d\zeta_i$. Since
\[
2 \zeta_i q_i \rho_i \varsigma \le \zeta_i^2  +  q_i^2 \varsigma^2, \qquad
2g_j(r - d  +  \tfrac12 \varsigma^2) \le  g_j^2  +  C(1  +  \varsigma^4),
\]
its generator also satisfies $\mathcal LG \le  CG$, and it does not explode. Let $\tau_n$ be the exit time of $\Zs$ from the ball of radius $n$. The stopped exponential $\mathcal E_{\cdot\wedge\tau_n}$ has a bounded integrand, so it defines a probability measure $\QQ_n$ under which $\Zs^{\tau_n}$ solves the share-measure system up to $\tau_n$. By uniqueness in law, $\QQ_n(\tau_n > T)$ equals the probability that the share-measure solution stays in the ball up to $T$, which tends to one. Hence
\[
\EE^{\QQ^{\Lambda}}[\mathcal E_T] \ge \EE^{\QQ^{\Lambda}} [\mathcal E_{T\wedge\tau_n} \mathbf 1_{\{\tau_n > T\}} ] = \QQ_n(\tau_n > T) \to  1.
\]
A positive local martingale is a supermartingale, so $\EE^{\QQ^{\Lambda}}[\mathcal E_T] = 1$, \cite{Sin1998}.
\end{proof}

The share measure adds the drift $\varsigma^2$ to every leverage state. For arithmetic returns the share-measure system has the same structure, and the argument again uses \cref{eq:growth_link}. The weaker condition of \cref{prop:arithmetic} gives existence of the state process but not the martingale property, so arithmetic returns would not relax the assumptions of the pricing results. In the one-mode family of \cref{thm:dichotomy}, under the lower bound on $U''$ in its part (i), the function $H$ from that proof also controls the share-measure system, in which the leverage drift of $w$ becomes $-\tfrac c2 \Phi$ for log returns and $-c \Phi$ for arithmetic returns. There the discounted index is a true martingale for $p \le  m-1$ under either convention, beyond the range $p \le  m/2$ of \cref{eq:growth_link}.

\begin{myremark} \label{rem:price_memory}
Price-side memory that acts only on the physical drift is compatible with the model, unlike the paired friction and noise excluded in \cref{rem:twod}. A term $-\int_0^tK_{XX}(t-s)\,dX_s$ with a kernel given by a finite sum of exponentials adds to $\mu$ a linear combination of exponentially weighted sums of past log returns. These sums can be carried as leverage states with $c_j\equiv0$, so the state keeps its form. Under every $\QQ^{\Lambda}$ the drift of $dS_t/S_t$ is $r-d$, so such memory acts only through the market price of traded risk $\theta$, that is, through the drifts of the noise modes and through the premium. These leverage states therefore remain part of the pricing state. Assumption (A5) requires $\theta$ to be bounded. A linear kernel can be saturated by adding to $\theta$ a bounded function of these states, which adds $\varsigma$ times that function to $\mu$. Since $\varsigma\ge\ell_{\min}f_{\min}$, a bounded addition to $\mu$ also keeps $\theta$ bounded. A linear kernel, as in the companion paper \cite{ItkinKazbekGLE}, makes $\theta$ grow linearly in them. The proof of \cref{thm:wellposed} still applies, because $\mu^2 \le C(1 + \varsigma^2 + \sum_jg_j^2) \le CG$. The martingale measures of \cref{prop:emm}, \cref{lem:martingale}, and the pricing results of this section use a bounded Sharpe ratio, however, and would need versions for an unbounded one.
\end{myremark}

\subsection{Selection by exponential utility} \label{sec:utility}

The premium is not fixed by no-arbitrage. We select it by exponential utility. Take a deterministic initial state $\mathbf z_0$, an investment horizon $T_H \le  T^\ast$ beyond every date on which a calibration payoff depends, and absolute risk aversion $\chi > 0$. The filtration is then generated by $\mathbf B$. We write $\QQ_0$ for the measure $\QQ^{\Lambda}$ of \cref{prop:emm} with $\boldsymbol\Lambda = 0$. Holding $\pi_t$ units of forward wealth in the total-return index gives the gains
\[
\mathcal G_t(\pi) = \int_0^t\pi_s(\bar\mu_s\,ds  +  \varsigma_s\,dW^X_s).
\]
Let $\mathcal M_f$ be the set of equivalent local martingale measures for $\tilde S$ with finite relative entropy $H(\QQ|\PP)$, and let $\mathcal A$ be the set of strategies for which $\mathcal G(\pi)$ is a $\QQ$-martingale for every $\QQ\in\mathcal M_f$, \cite{DelbaenEtAl2002}. The investor maximizes
\[
\EE^{\PP} [ -\exp(-\chi( \Pi_0  +  \mathcal G_{T_H}(\pi))) ]
\]
over $\mathcal A$.

For $\mathbf v\in\mathbb R^{N  +  1}$ write $\mathbf v^\perp = \mathbf v - (\mathbf v\cdot\mathbf r)\mathbf r$ for its component orthogonal to $\mathbf r$. The selected measure is described by the BSDE
\begin{equation} \label{eq:bsde}
\Psi_t = \int_t^{T_H}\Big(\tfrac12|\mathbf V_s^\perp|^2 - \theta_s\,\mathbf V_s\cdot\mathbf r - \tfrac12\theta_s^2\Big)\,ds - \int_t^{T_H}\mathbf V_s\cdot d\mathbf B_s, \qquad t\in[0,T_H],
\end{equation}
for a real process $\Psi$ and an $\mathbb R^{N  +  1}$-valued process $\mathbf V$. Its driver has quadratic growth in $\mathbf V$, and its terminal value is zero. The next theorem specializes known results on exponential utility in a Brownian filtration, \cite{DelbaenEtAl2002, HuImkellerMuller2005, ManiaSchweizer2005} to the present market. We give the proof for completeness, because the market price of traded risk depends on the lifted state and is controlled only through (A5).

\begin{theorem}[The utility-selected measure] \label{thm:memm}
Under \cref{ass:std}:
\begin{enumerate}
\item[(a)] \Cref{eq:bsde} has a unique solution $(\Psi,\mathbf V)$ with $\Psi$ bounded. For it, $\int\mathbf V\cdot d\mathbf B$ is a martingale of bounded mean oscillation (BMO), and $-\tfrac12\bar\theta^2(T_H - t) \le \Psi_t \le  0$.

\item[(b)] Put $\boldsymbol\Lambda^0 = \mathbf V^\perp$ and $\pi^\ast = (\theta  +  \mathbf V\cdot\mathbf r)/(\chi\varsigma)$. Then $\mathcal E(\int(-\theta\mathbf r  +  \boldsymbol\Lambda^0)\cdot d\mathbf B)$ is a uniformly integrable martingale, and the measure $\QQ^{(0)}$ it defines belongs to $\mathcal M_f$, with
\[
d\QQ^{(0)}/d\PP = \exp\big(-\Psi_0 - \chi\mathcal G_{T_H}(\pi^\ast)\big), \qquad H(\QQ^{(0)}|\PP) = -\Psi_0.
\]

\item[(c)] $\pi^\ast\in\mathcal A$ is optimal, with value $-\exp(-\chi\Pi_0  +  \Psi_0)$.

\item[(d)] $\QQ^{(0)}$ is the minimal entropy martingale measure: $H(\QQ|\PP) = H(\QQ|\QQ^{(0)})  +  H(\QQ^{(0)}|\PP)$ for every $\QQ\in\mathcal M_f$.

\item[(e)] The marginal utility-based price, \cite{Davis1997}, of a bounded $\mathcal F_T$-measurable claim $H$ with $T \le  T_H$ is $\EE^{\QQ^{(0)}}[e^{-\int_0^Tr_s\,ds}H]$.
\end{enumerate}
\end{theorem}

\begin{proof}
See \cref{app:theorem5}.
\end{proof}

For a function $\omega(t,\mathbf z)$ of class $C^{1,2}$, define its exposures $v_i(\omega) = q_i\partial_{\zeta_i}\omega$ and $e(\omega) = \varsigma\,(\partial_x\omega  +  \sum_j\partial_{g_j}\omega)$, the exposure vector $\mathbf V(\omega) = e(\omega)\mathbf r  +  (v_1,\ldots,v_N,0)$, and
\begin{equation} \label{eq:residual_variance}
\mathcal R(\omega) = |\mathbf V(\omega)^\perp|^2 = \sum_iv_i(\omega)^2 - \Big(\sum_i\rho_iv_i(\omega)\Big)^2\ge 0.
\end{equation}
The leverage states contribute only to the traded exposure $e(\omega)$, in agreement with \cref{rem:leverage_entry}. Let $\mathcal L^{\QQ_0}$ be the generator of \cref{eq:lift_S} with $\boldsymbol\Lambda = 0$, and consider
\begin{equation} \label{eq:ptoq_omega}
\partial_t\omega  +  \mathcal L^{\QQ_0}\omega  +  \tfrac12\mathcal R(\omega) - \tfrac12\theta^2 = 0, \qquad \omega(T_H,\cdot) = 0.
\end{equation}

\begin{theorem}[Feedback form of the premium] \label{prop:feedback}
Assume, in addition to \cref{ass:std}, that \cref{eq:ptoq_omega} admits a bounded classical solution $\omega\in C^{1,2}([0,T_H)\times\mathbb R^{2  +  2N  +  M})\cap C([0,T_H]\times\mathbb R^{2  +  2N  +  M})$ with $\mathbf V(\omega)$ bounded on $[0,T_H)\times\mathbb R^{2 + 2N + M}$, and that this solution depends on $(u,\zeta)$ only through $h = \zeta - u$ (equivalently, $\omega(t,\mathbf z) = \tilde\omega(t,P_{\mathrm r}(\mathbf z))$ for a function $\tilde\omega$ on the reduced state, where $P_{\mathrm r}(\mathbf z) = (x,y,\zeta - u,g)$). Then $\Psi_t = \omega(t,\Zs_t)$ and $\mathbf V_t = \mathbf V(\omega)(t,\Zs_t)$, so the premium of \cref{thm:memm} has the feedback form
\begin{equation} \label{eq:ptoq_mpr}
\Lambda_i^0 = v_i(\omega) - \rho_i\sum_k\rho_kv_k(\omega), \qquad \Lambda_0^0 = -\sqrt{1 - \rho^2}\sum_k\rho_kv_k(\omega),
\end{equation}
evaluated at $(t,\Zs_t)$. The premium then depends on $(u,\zeta)$ only through $h$. If it is also locally Lipschitz in $\mathbf z$, uniformly in $t$, then \cref{lem:martingale} applies to $\QQ^{(0)}$. In general, the existence of such a classical solution, and its dependence on $(u,\zeta)$ only through $h$, are additional regularity hypotheses.
\end{theorem}

\begin{proof}
Under $\QQ_0$, $d\mathbf B = d\mathbf B^{\QQ_0} - \theta\mathbf r\,dt$, so It\^o's formula and \cref{eq:ptoq_omega} give, under $\PP$,
\[
d\omega(t,\Zs_t) = \big[\tfrac12\theta^2 - \tfrac12|\mathbf V(\omega)^\perp|^2  +  \theta\,\mathbf V(\omega)\cdot\mathbf r\big]\,dt  +  \mathbf V(\omega)\cdot d\mathbf B.
\]
Hence $(\omega(\cdot,\Zs),\mathbf V(\omega)(\cdot,\Zs))$ solves \cref{eq:bsde} with a bounded first component, and uniqueness in \cref{thm:memm}(a) gives the identification. The components of $\mathbf V(\omega)^\perp$ are \cref{eq:ptoq_mpr}.
\end{proof}

The coefficient on $B_0$ is needed for orthogonality to $\mathbf r$, so the one-factor expression of a white-noise model does not apply when several noise modes are present.

\begin{corollary}[Deterministic Sharpe ratio] \label{cor:deterministic_sharpe}
If $\theta_t=\theta(t)$ is bounded, continuous and deterministic, then
\begin{equation} \label{eq:deterministic_sharpe}
\Psi_t=-\tfrac12\int_t^{T_H}\theta(s)^2\,ds, \qquad \mathbf V=0, \qquad \boldsymbol\Lambda^0=0,
\end{equation}
and the selected measure is $\QQ_0$, with density $\mathcal E(-\int_0^\cdot\theta(s)\,dW^X_s)_{T_H}$. The function $\omega(t,\mathbf z) = -\tfrac12\int_t^{T_H}\theta(s)^2\,ds$ solves \cref{eq:ptoq_omega} and satisfies the hypotheses of \cref{prop:feedback}.
\end{corollary}

\begin{proof}
The pair in \cref{eq:deterministic_sharpe} is bounded and solves \cref{eq:bsde}, and uniqueness in \cref{thm:memm}(a) applies. All state derivatives of $\omega$ vanish, so $\mathbf V(\omega) = 0$ and $\mathcal R(\omega) = 0$, while $\partial_t\omega = \theta(t)^2/2$.
\end{proof}

A nonconstant state-dependent Sharpe ratio gives, in general, a nonzero premium. \Cref{thm:memm} provides it as a BMO process from the standing assumptions. Its feedback form and boundedness, which \cref{lem:martingale} and the reduced Markov representation under $\QQ^{(0)}$ use, require a classical solution of \cref{eq:ptoq_omega} with bounded exposures. The next theorem gives such a solution without leverage. It uses the distortion transform of \cite{Zariphopoulou2001}, which linearizes the semilinear equation when the residual term is a constant multiple of the second-order part of the generator.

\begin{theorem}[Classical solution without leverage] \label{prop:noleverage}

In addition to \cref{ass:std}, assume that $c_j\equiv0$ for all $j$, and that either $N=1$ or $\rho_1=\cdots=\rho_N=0$. Put $\kappa_\rho=1-\rho^2$. Let the Sharpe ratio have the form $\theta(t,\mathbf z)=\tilde\theta(t,y,\zeta-u)$, where $\tilde\theta$ and its first and second derivatives in $(y,h)$ are continuous in $(t,y,h)$ and bounded. Let $U\in C^3(\mathbb R)$ with $U''\ge-K$ and $|U'''| \le  K(1 + U-U_\ast)$ for a constant $K$. Then
\begin{equation} \label{eq:distortion}
\omega(t,\mathbf z)=\frac{1}{\kappa_\rho}\log\EE^{\QQ_0}_{t,\mathbf z}\Big[\exp\Big(-\frac{\kappa_\rho}{2}\int_t^{T_H}\theta(s,\Zs_s)^2\,ds\Big)\Big],
\end{equation}
where $\EE^{\QQ_0}_{t,\mathbf z}$ refers to the solution of \cref{eq:lift_S} with $\boldsymbol\Lambda=0$ started from $\mathbf z$ at time $t$, is a solution of \cref{eq:ptoq_omega} in $C^{1,2}([0,T_H)\times\mathbb R^{2 + 2N + M})\cap C([0,T_H]\times\mathbb R^{2 + 2N + M})$, and $-\tfrac12\bar\theta^2(T_H-t) \le \omega \le 0$. It has the form $\omega(t,\mathbf z)=\tilde\omega(t,y,\zeta-u)$, and its exposures $e(\omega)=0$ and $v_i(\omega)=q_i\partial_{h_i}\tilde\omega$ are bounded. The hypotheses of \cref{prop:feedback} therefore hold, the premium \cref{eq:ptoq_mpr} is bounded and locally Lipschitz in $\mathbf z$ uniformly in $t$, and \cref{lem:martingale} applies to $\QQ^{(0)}$.
\end{theorem}

\begin{proof}
See \cref{app:theorem7}.
\end{proof}

The conditions on $U$ hold for quadratic potentials, for cubic--quartic potentials with $c_4>0$, including double wells, and for two-sided exponential potentials such as $\cosh(\alpha y)$. Unlike that of \cref{cor:deterministic_sharpe}, the premium of \cref{prop:noleverage} is in general not zero. The transform requires the residual term to be a constant multiple of the second-order part of the generator acting on $\omega$. With leverage, the variance driver depends on the leverage states, so $\omega$ depends on the leverage states $g_j$, which carry the noise of the traded index, and, through $\varsigma$, on $x$. Its traded exposure $e(\omega)$ is then in general nonzero, the second-order part of the generator includes the traded direction, and $\mathcal R(\omega)$ covers only the unhedgeable part, so the transform no longer removes the quadratic term.

The same obstruction appears with several modes correlated with returns, because $\sum_i v_i^2 - (\sum_i\rho_iv_i)^2$ is not a constant multiple of $\sum_iv_i^2$. It also appears when the Sharpe ratio depends on leverage states, as for the price-side memory of \cref{rem:price_memory}. For stochastic volatility models with a one-dimensional volatility factor, \cite{BenthKarlsen2005} express the density of the minimal entropy measure through a semilinear equation of this type and apply it to the Stein--Stein and Heston models.

The obstruction disappears when the Sharpe ratio depends on the state only through a combination of state variables that carries no traded noise and evolves autonomously and its explicit return increments cancel.. With constant leverage this happens in the following tuned family, in which two leverage modes with rates on either side of an intrinsic relaxation rate cancel the return noise.

\begin{proposition}[Classical solution with constant leverage] \label{prop:ou_leverage}
In addition to \cref{ass:std}, let $N=1$, with parameters $a,\gamma,q>0$ and correlation $\rho_1=\rho$, and let $U(y)=\tfrac12a_Uy^2$ with $a_U>0$. Let $\nu_\star$ be the smaller root of
\[
\nu^2 - (a_U + a + \gamma) \nu + a_U \gamma = 0,
\]
so that $0<\nu_\star<\min\{a_U,\gamma\}$. Let $M=2$, with rates $0<\lambda_1<\nu_\star<\lambda_2$ and constant amplitudes $c_1=\ell_c(\nu_\star-\lambda_1)$ and $c_2=\ell_c(\lambda_2-\nu_\star)$ for some $\ell_c>0$. Put
\[
\iota(\mathbf z)=\frac{a_U}{\nu_\star}\,y + \frac{\zeta-u}{\gamma-\nu_\star} + \ell_c\,(g_1-g_2), \qquad \sigma_\iota=\frac{q}{\gamma-\nu_\star}, \qquad \kappa_\rho=1-\rho^2,
\]
and let the Sharpe ratio be $\theta(t,\mathbf z)=\vartheta(\iota(\mathbf z))$, where $\vartheta$ is bounded with bounded derivatives up to order three. Then $\iota_t=\iota(\Zs_t)$ satisfies
\begin{equation} \label{eq:ou_factor}
d\iota_t=-\nu_\star\iota_t\,dt + \sigma_\iota\,dB_{1,t}\ \mathrm{under }\PP, \qquad d\iota_t=-\big(\nu_\star\iota_t + \rho\sigma_\iota\vartheta(\iota_t)\big)\,dt + \sigma_\iota\,dB^{\QQ_0}_{1,t}\ \mathrm{under }\QQ_0.
\end{equation}

Let
\[
\Omega(t,\iota) = \kappa_\rho^{-1} \log \EE^{\QQ_0} \big[ \exp \big(-\tfrac{ \kappa_\rho}{2} \int_t^{T_H} \vartheta(\iota_s)^2\,ds \big)\, \big|\, \iota_t = \iota \big].
\]
Then $\omega(t,\mathbf z)=\Omega(t,\iota(\mathbf z))$ is a bounded classical solution of \cref{eq:ptoq_omega} with bounded exposures, and it depends on $(u,\zeta)$ only through $h$. Its premium, $\Lambda^0_1 = \kappa_\rho \sigma_\iota\, \partial_\iota \Omega$ and  $\Lambda^0_0 = - \rho \sqrt{\kappa_\rho}\, \sigma_\iota\, \partial_\iota \Omega$ at $(t,\iota(\Zs_t))$, is bounded and globally Lipschitz in $\mathbf z$ uniformly in $t$, and \cref{lem:martingale} applies to $\QQ^{(0)}$.
\end{proposition}

\begin{proof}
Write $\tilde m=-a_Uy + h-c_1g_1-c_2g_2$ for the drift of $Y$. The root equation gives $a_U/\nu_\star-a/(\gamma-\nu_\star)=1$. The return increments cancel in $dg_1-dg_2=-(\lambda_1g_1-\lambda_2g_2)\,dt$, so \cref{prop:reduced} gives, under $\PP$,
\[
d\iota_t=\Big(\tilde m-\frac{\gamma h_t}{\gamma-\nu_\star}-\ell_c\lambda_1g_{1,t} + \ell_c\lambda_2g_{2,t}\Big)\,dt + \sigma_\iota\,dB_{1,t}=-\nu_\star\iota_t\,dt + \sigma_\iota\,dB_{1,t},
\]
where the second equality uses the root identity and the choice of $c_1$ and $c_2$. Under $\QQ_0$, $dB_{1,t}=dB^{\QQ_0}_{1,t}-\rho\theta_t\,dt$, which gives \cref{eq:ou_factor}. The drift $m_\iota(\iota)=-\nu_\star\iota-\rho\sigma_\iota\vartheta(\iota)$ of this scalar diffusion has bounded first and second derivatives, so its flow has first and second derivatives bounded by deterministic constants on $[t,T_H]$. As in the proof of \cref{prop:noleverage}, $\tilde\psi=e^{\kappa_\rho\Omega}$ is therefore a classical solution of
\[
\partial_t\tilde\psi + m_\iota\,\partial_\iota\tilde\psi + \tfrac12\sigma_\iota^2\,\partial_\iota^2\tilde\psi-\tfrac{\kappa_\rho}{2}\vartheta^2\tilde\psi=0, \qquad \tilde\psi(T_H,\cdot)=1,
\]
with $e^{-\kappa_\rho\bar\theta^2(T_H-t)/2}\le\tilde\psi\le1$, and $\Omega$, $\partial_\iota\Omega$ and $\partial_\iota^2\Omega$ are bounded on $[0,T_H)\times\mathbb R$. Since $\iota$ is linear in $\mathbf z$ with $(\partial_x + \partial_{g_1} + \partial_{g_2})\iota=0$, the traded exposure of $\omega$ vanishes, its intrinsic exposure is $v_1(\omega)=\sigma_\iota\partial_\iota\Omega$, and $\mathcal R(\omega)=\kappa_\rho\sigma_\iota^2(\partial_\iota\Omega)^2$. The quadratic variation of $\iota$ is $\sigma_\iota^2\,dt$, so \cref{eq:ptoq_omega} for $\omega$ reduces to the equation for $\Omega$ that the transform of \cref{prop:noleverage} turns into the linear equation for $\tilde\psi$. The premium follows from \cref{eq:ptoq_mpr}, and its bounds follow from those on $\partial_\iota\Omega$ and $\partial_\iota^2\Omega$.
\end{proof}

\Cref{prop:ou_leverage} retains positive leverage amplitudes, a nonconstant local function and a nonconstant volatility map, and the Sharpe function $\vartheta$ is free. For example, with $a_U=a=\gamma=1$, so that $\nu_\star=(3-\sqrt5)/2$, with $\lambda_1=\nu_\star/2$, $\lambda_2=2\nu_\star$ and $\vartheta(\iota)=\theta_0 + \varepsilon_\vartheta\tanh\iota$ for $\theta_0>|\varepsilon_\vartheta|>0$, the premium is not zero, since $\partial_\iota\Omega(T_H-\tau,0)=-\tau\theta_0\varepsilon_\vartheta + o(\tau)$. The family is tuned. It requires a quadratic potential, one intrinsic mode, constant amplitudes, and the stated relation between amplitudes and rates. For general Sharpe ratios with leverage, whether \cref{eq:ptoq_omega} has a bounded classical solution with bounded exposures remains open. Hypoellipticity of the state process does not by itself provide the global gradient bounds needed for such a solution. The traded-risk correction $-q_i\rho_i\theta$ changes the intrinsic modes under pricing even when the residual premium vanishes.

Neither \cref{eq:bsde} nor \cref{eq:ptoq_omega} contains the risk aversion, so
the baseline measure carries no free risk-aversion parameter. The zero-position
limit of exponential indifference prices also equals the minimal entropy price,
\cite{Becherer2003}. All calibration instruments are priced linearly under the
one measure $\QQ^{(0)}$. This preserves static aggregation and, in particular,
the replication of the log contract by a strip of SPX options.

An inventory-based premium is obtained by fixing a reference liability $H^{\mathrm{ref}}$ at $T_H$ held short in quantity $\nu > 0$, with $\EE^{\QQ^{(0)}}[\exp(\chi\nu H^{\mathrm{ref}})] < \infty$. Bounded claims satisfy this. For unbounded claims such as variance swaps it depends on the tails of the variance, and it fails when the variance has lognormal tails. The certainty equivalent of the position exceeds that of the investment problem by $\nu p$, where the per-unit price solves $\partial_tp  +  \mathcal L^{\QQ^{(0)}}p  +  \tfrac12\chi\nu\mathcal R(p) = 0$ with $p(T_H,\cdot) = H^{\mathrm{ref}}$. The marginal measure around this position has premium $\boldsymbol\Lambda^\nu = \boldsymbol\Lambda^0  +  \chi\nu\,\mathbf V(p)^\perp$, and only the product $\chi\nu$ enters. For bounded $H^{\mathrm{ref}}$ the corresponding quadratic BSDE has a bounded solution by the argument of \cref{thm:memm}, and the feedback form requires the regularity of \cref{prop:feedback}.

\subsection{The reduced state under the pricing measure} \label{sec:reduced_Q}

\begin{proposition}[Reduction under the selected measure] \label{prop:reduced_Q}
Assume that $\mu$ depends on $(u,\zeta)$ only through $h = \zeta - u$, and let $\tilde\omega$ be a reduced-state solution of \cref{eq:ptoq_omega} satisfying the hypotheses of \cref{prop:feedback}. Then $\omega = \tilde\omega\circ P_{\mathrm r}$ satisfies the hypotheses of \cref{prop:feedback} on the full state. The premium $\boldsymbol\Lambda^0$ then depends on $(u,\zeta)$ only through $h$, and the reduced state is Markov under $\QQ^{(0)}$, with
\begin{equation} \label{eq:reduced_lift_Q}
dh_{i,t} = -\big(\gamma_ih_{i,t}  +  a_i\tilde m(\Zs^{\mathrm r}_t)  +  q_i\rho_i\theta_t - q_i\tilde\Lambda^0_{i,t}\big)\,dt  +  q_i\,dB^{\QQ}_{i,t}.
\end{equation}
\end{proposition}

\begin{proof}
By the chain rule, $\partial_{\zeta_i}\omega = \partial_{h_i}\tilde\omega$ and $\partial_{u_i}\omega = -\partial_{h_i}\tilde\omega$, with the other first derivatives unchanged. The first-order part of $\mathcal L^{\QQ_0}\omega$ therefore contains $\partial_{h_i}\tilde\omega$ multiplied by the difference of the drifts of $\zeta_i$ and $u_i$, which is the drift of $h_i$. The coordinates $u_i$ carry no diffusion, so the second-order parts coincide, and $\mathcal L^{\QQ_0}\omega = (\tilde{\mathcal L}^{\QQ_0}\tilde\omega)\circ P_{\mathrm r}$. The exposures $v_i$ and $e$ involve only $\partial_{\zeta_i}$, $\partial_x$, and $\partial_{g_j}$, so $\mathbf V(\omega) = \tilde{\mathbf V}(\tilde\omega)\circ P_{\mathrm r}$ and $\mathcal R(\omega) = \tilde{\mathcal R}(\tilde\omega)\circ P_{\mathrm r}$. Hence $\omega$ solves \cref{eq:ptoq_omega}, and the remaining claims follow from \cref{prop:feedback} and as in \cref{prop:reduced}.
\end{proof}

The premium changes only the drifts of the coordinates $h_i$, which are spanned by the diffusion fields. The proof of \cref{prop:hypo} therefore applies without change under $\QQ^{\Lambda}$ when $\boldsymbol\Lambda$ and $t\mapsto r_t - d_t$ are $C^\infty$. Thus, if the hypotheses of \cref{prop:hypo} hold and the selected premium is $C^\infty$, the reduced state under $\QQ^{(0)}$ has a smooth density solving the forward equation. \Cref{prop:feedback} alone assumes less regularity and does not imply this additional smoothness. The reduction requires the premium to depend on $(u,\zeta)$ only through $h$, which is automatic when the premium has the feedback form of \cref{prop:feedback} and $\tilde\omega$ is a function of the reduced state.

\section{The local volatility function} \label{sec:localvol}

Throughout this section $\QQ$ denotes a measure $\QQ^{\Lambda}$ as in \cref{lem:martingale}, for example $\QQ^{(0)}$ when the hypotheses of \cref{prop:feedback} hold, and $\EE = \EE^{\QQ}$.

\subsection{Markovian projection} \label{sec:leverage_projection}

\begin{corollary}[Projection] \label{prop:projection}
For every $t\in[0,T^\ast]$, the law of $S_t$ under $\QQ$ coincides with the law at time $t$ of a weak solution of
\begin{equation} \label{eq:gyongy}
d\hat S_t/\hat S_t = (r_t - d_t)\,dt  +  \Dupvol(t,\hat S_t)\,d\hat W_t, \qquad
\Dupvol^2(t,K) = L^2(t,K)\,\EE\big[F^2(Y_t)\,\big|\,S_t = K \big].
\end{equation}
\end{corollary}

\begin{proof}
Apply the mimicking theorem of \cite{BrunickShreve2013} to $X = \log S$, whose drift and squared diffusion coefficient are $r - d - \tfrac12\varsigma^2$ and $\varsigma^2$. Its integrability condition,
\[
\EE\int_0^{T^\ast}(|r - d - \tfrac12\varsigma_t^2|  +  \varsigma_t^2)\,dt < \infty,
\]
holds because $\varsigma^2 \le  CG$ and by \cref{lem:martingale}. The conditional expectations of the coefficients given $X_t$ are $r - d - \tfrac12\Dupvol^2$ and $\Dupvol^2$, by \cref{eq:gyongy}, and It\^o's formula transfers the result to $\hat S = e^{\hat X}$.
\end{proof}

Exact calibration to an arbitrage-free surface with Dupire local volatility $\Dupvol$, \cite{Dupire1994}, therefore requires $L^2(t,K) = \Dupvol^2(t,K) / \EE[F^2(Y_t)| S_t = K]$. The denominator depends on the law of the solution, which depends on $L$, so exact calibration is a fixed point of McKean--Vlasov type. Existence is known only in special cases, \cite{LackerShkolnikovZhang2020}, and the fixed-point iteration of \cite{GuyonHenryLabordere2011} is a numerical method rather than an existence result. The local function is the multiplicative correction that reconciles the GLE with the surface. The choice $L\equiv1$, when contained in the admissible class, gives an exact fit precisely when the GLE alone reproduces the surface.

Since $F\ge f_{\min}$, the denominator is at least $f_{\min}^2$. An exact calibration to a surface with $\Dupvol \le \sigma_{\max}$ therefore satisfies $L \le \sigma_{\max}/f_{\min}$. No lower bound on $L$ follows without an upper bound on the conditional variance, which in the wings depends on the tails of the driver.

The conditional expectation in \cref{eq:gyongy} requires only the joint law of a finite-dimensional Markov process. If the reduced state of \cref{prop:reduced_Q} has a density $p_t$, then
\begin{equation} \label{eq:condexp_lift}
\EE\big[F^2(Y_t)\,\big|\,S_t = K\big] = \frac{\int F^2(y)\,p_t(\log K,y,h,g)\,dy\,dh\,dg}{\int p_t(\log K,y,h,g)\,dy\,dh\,dg}
\end{equation}
wherever the denominator is positive. The density, when it exists and is smooth, solves the forward equation of the reduced generator. A sufficient condition is the H\"ormander bracket condition verified in \cref{prop:hypo}, with the pricing extension and its additional smoothness hypotheses given in \cref{sec:reduced_Q}. Finite dimensionality alone does not imply the existence of a smooth density. The same expectation can be estimated by simulating the reduced state.

\subsection{The VIX} \label{sec:vix}

Let $\Delta$ be thirty calendar days in years. VIX is expressed in decimal annualized volatility units. Quotation in percentage points multiplies it by $100$. Following the log-contract definition, \cite{CarrMadan1998, DemeterfiEtAl1999}, set
\begin{equation} \label{eq:vix_log_contract}
\mathrm{VIX}_T^2 = -\frac{2}{\Delta}\EE\Big[\log\frac{S_{T  +  \Delta}}{S_T} - \int_T^{T  +  \Delta}(r_u - d_u)\,du\,\Big|\,\mathcal F_T\Big].
\end{equation}

\begin{corollary}[VIX representation] \label{prop:vix}
For $T  +  \Delta \le  T^\ast$,
\begin{equation} \label{eq:vixdef}
\mathrm{VIX}_T^2 = \frac1\Delta\EE\Big[\int_T^{T  +  \Delta}L^2(u,S_u)F^2(Y_u)\,du\,\Big|\,\mathcal F_T\Big] = \frac1\Delta\mathcal V_T(\Zs_T),
\end{equation}
where $\mathcal V_T(\mathbf z) = \EE_{T,\mathbf z}[\int_T^{T  +  \Delta} \varsigma_u^2\,du]$ is a Borel function. If $w\in C^{1,2}([T,T  +  \Delta)\times\mathbb R^{2  +  2N  +  M})$ solves $\partial_tw  +  \mathcal L^{\QQ}w + \varsigma^2 = 0$ with $w(T  +  \Delta,\cdot) = 0$, and $w(t,\Zs_t)$, $t\in[T,T  +  \Delta]$, is of class (D), then $w(T,\cdot) = \mathcal V_T$ on the support of the law of $\Zs_T$.
\end{corollary}

\begin{proof}
By \cref{eq:lift_S},
\[
\log(S_{T  +  \Delta}/S_T) = \int_T^{T  +  \Delta}(r_u - d_u - \tfrac12 \varsigma_u^2)\,du  +  \int_T^{T  +  \Delta} \varsigma_u\,dB^{\QQ}_u.
\]
The stochastic integral has zero conditional expectation because $\EE\int_0^{T^\ast}\varsigma_u^2\,du < \infty$. This gives the first identity. The second follows from the Markov property of \cref{lem:martingale}. For the last claim, It\^o's formula shows that $w(t,\Zs_t)  +  \int_T^t\varsigma_u^2\,du$ is a local martingale on $[T,T  +  \Delta]$. It is of class (D), hence a martingale, and its terminal value is $\int_T^{T  +  \Delta}\varsigma_u^2\,du$.
\end{proof}

The integrand is the full local stochastic variance $L^2F^2$, not $V$ alone. The local function therefore enters both the SPX marginals and the quantity averaged to construct the VIX, and the two markets cannot be fitted separately. Because the log contract is replicated by a strip of SPX options and all prices are linear under $\QQ$, \cref{eq:vix_log_contract} also ties the model VIX to the model SPX surface.

The identity \cref{eq:vixdef} defines the model VIX as the continuous log contract. The CBOE VIX is computed from a finite strip of options with a truncated strike range and a discrete Riemann sum, and it differs from the continuous log contract by the truncation and discretization error of that formula. This error is a well-studied source of discrepancy, independent of the model. Model prices of VIX derivatives computed from \cref{eq:vixdef} neglect this difference, which is small but not negligible.

\begin{myremark} \label{rem:rho_vix}
In an autonomous stochastic volatility model the law of the VIX does not depend on the return--variance correlation. Here this conclusion requires three conditions: $L$ does not depend on $S$, the leverage amplitudes vanish, and the pricing drifts of the noise modes do not depend on $\rho$. Each can fail. A spot-dependent $L$ brings the joint law of spot and driver into \cref{eq:vixdef}, the leverage term couples the driver to returns, and the drift $-q_i\rho_i\theta$ in \cref{eq:lift_S} depends on $\rho$ whenever $\bar\mu\ne 0$.
\end{myremark}

The leverage feedback and the local function change VIX prices through the law of the state at expiry and the forward variance at expiry. A first-order expansion separates them. Let $\mathcal L_\varepsilon$ be the pricing generator of the reduced state for a one-parameter family of model inputs, with $\mathcal L_\varepsilon = \mathcal L_0  +  \varepsilon\mathcal A  +  o(\varepsilon)$, let $\mathcal V_{T,\varepsilon}$ be the forward variance of \cref{prop:vix}, and let $C_\varepsilon(t,\mathbf z)$ be the price at time $t \le  T$ of a payoff $H(\mathrm{VIX}_T)$ with $H\in C^1$ and $H'$ bounded.

\begin{proposition}[First-order sensitivity] \label{prop:rho_dependent}
Assume that $\mathcal V_{T,\varepsilon}$ and $C_\varepsilon$ are of class $C^{1,2}$, differentiable in $\varepsilon$ at $0$ with derivatives $\dot{\mathcal V}_T$ and $\dot C$ of class $C^{1,2}$, that the processes obtained by evaluating these functions and their images under $\mathcal A$ along the $\varepsilon = 0$ dynamics are of class (D), and that $\mathcal V_{T,0}(T,\cdot) > 0$. Then, with $v_{T,0} = \sqrt{\mathcal V_{T,0}(T,\cdot)/\Delta}$,
\begin{align}
\dot{\mathcal V}_T(T,\mathbf z) &= \EE^{0}_{T,\mathbf z}\Big[\int_T^{T  +  \Delta}\big(\mathcal A\mathcal V_{T,0}  +  \dot{\mathfrak a}\big)(u,\Zs^{\mathrm r}_u)\,du\Big], \label{eq:feedback_forward_representation} \\
\dot C(t,\mathbf z) &= \EE^{0}_{t,\mathbf z}\Big[e^{-\int_t^Tr_s\,ds}H'(v_{T,0})\frac{\dot{\mathcal V}_T(T,\Zs^{\mathrm r}_T)}{2\Delta v_{T,0}}  +  \int_t^Te^{-\int_t^ur_s\,ds}(\mathcal AC_0)(u,\Zs^{\mathrm r}_u)\,du\Big]. \label{eq:vix_price_derivative}
\end{align}
where $\EE^0$ refers to the $\varepsilon = 0$ dynamics, $\mathfrak a_\varepsilon = \varsigma_\varepsilon^2$ is the instantaneous variance of the family, and $\dot{\mathfrak a}$ is its derivative at $\varepsilon = 0$. For leverage amplitudes $c_j^\varepsilon = \varepsilon c_j$, with given amplitude functions $c_j$ and with $L$, $F$ and the intrinsic parameters fixed, use the right derivative at $\varepsilon=0$. Then $\dot{\mathfrak a} = 0$ and, for a smooth test function $\varphi$,
\begin{equation} \label{eq:feedback_generator_derivative}
\mathcal A\varphi = -\Big(\sum_jc_j(y)g_j\Big)\Big(\partial_y\varphi - \sum_ia_i\partial_{h_i}\varphi\Big)  +  \sum_iq_i(\dot\Lambda_i-\rho_i\dot\theta)\partial_{h_i}\varphi,
\end{equation}
with $\dot\Lambda_i$ and $\dot\theta$ the derivatives of the residual and traded-risk premiums at a fixed state. If the physical drift function is also held fixed, then $\dot\theta=0$. For a local function $L_\varepsilon = L_0e^{\varepsilon\ell}$, $\dot{\mathfrak a} = 2\ell\mathfrak a_0$, and $\mathcal A$ collects the derivative of the coefficients that contain $\varsigma$ and of the premium.
\end{proposition}

\begin{proof}
Differentiate
\[
\partial_t\mathcal V_{T,\varepsilon}  +  \mathcal L_\varepsilon\mathcal V_{T,\varepsilon}  +  \mathfrak a_\varepsilon = 0, \qquad \mathcal V_{T,\varepsilon}(T  +  \Delta,\cdot) = 0,
\]
at $\varepsilon = 0$. The derivative solves $\partial_t\dot{\mathcal V}_T  +  \mathcal L_0\dot{\mathcal V}_T  +  \mathcal A\mathcal V_{T,0}  +  \dot{\mathfrak a} = 0$ with zero terminal value. By It\^o's formula under the $\varepsilon = 0$ dynamics and the class (D) hypothesis, the process
\[
\dot{\mathcal V}_T(t,\Zs^{\mathrm r}_t)  +  \int_T^t(\mathcal A\mathcal V_{T,0}  +  \dot{\mathfrak a})\,du
\]
is a martingale, which gives \cref{eq:feedback_forward_representation}. The price solves $\partial_tC_\varepsilon +  \mathcal L_\varepsilon C_\varepsilon - r C_\varepsilon = 0$ with terminal value $H( \sqrt{\mathcal V_{T,\varepsilon}(T,\cdot )/\Delta})$. Its derivative solves the same equation with source $\mathcal AC_0$ and terminal value
\[
H'(v_{T,0}) \dot{\mathcal V}_T (T,\cdot) / (2\Delta v_{T,0}),
\]
and the same argument gives \cref{eq:vix_price_derivative}. Formula \cref{eq:feedback_generator_derivative} is the derivative of the drifts of $Y$ and $h_i$ in \cref{eq:reduced_lift_Q}.
\end{proof}

The first term in \cref{eq:vix_price_derivative} is the change of the VIX map at a given state, and the second is the change of the law of the state at expiry. With the leverage amplitudes set to zero, the baseline derivative $\partial_\rho C(\rho,0)$ is not zero in general, for the reasons in \cref{rem:rho_vix}, so state-dependent leverage modifies, rather than creates, the dependence of VIX prices on correlation.

\subsection{Calibration to bid--ask spreads} \label{sec:leverage_band}

Exact calibration is also ill conditioned in practice. The Dupire variance involves strike and maturity derivatives of a surface inferred from discrete noisy quotes, and the conditional expectation in \cref{eq:gyongy} is poorly determined where the spot marginal is small. Both amplify observation and numerical error, which motivates a tolerance on the quotes.

We replace exact calibration by a tolerance band and show that the resulting problem has a solution. Let $\mathcal C$ be a finite set of SPX calls with maturities $T_k$, strikes $K_k$, and quotes $C_k^{\mathrm{bid}} \le  C_k^{\mathrm{ask}}$, and let $\mathcal I$ be a finite set of VIX futures and options with maturities $T_m$, $T_m +  \Delta \le  T^\ast$, payoffs $H_m(\mathrm{VIX}_{T_m})$ that are Lipschitz, and market prices $P_m^{\mathrm{mkt}}$. Model prices are
\[
C_k(L,\vartheta) = \EE[e^{-\int_0^{T_k}r_u\,du}(S_{T_k} - K_k)^ + ], \qquad P_m(L,\vartheta) = \EE[e^{-\int_0^{T_m}r_u\,du}H_m(\mathrm{VIX}_{T_m})],
\]
with undiscounted expectations for futures.

The structural parameters $\vartheta$ range over a compact set $\mathcal K\subset\mathbb R^p$. They comprise the kernel exponents and amplitudes, the potential, the leverage amplitudes and sensitivity, $\rho$, and $\Theta$. The local functions range over
\begin{equation} \label{eq:admissible_L}
\mathfrak L = \big\{L:\ \ell_{\min} \le  L \le \ell_{\max},\ |\Lh(t,x) - \Lh(s,x')| \le  K_L\big(|t - s|^\alpha  +  |x - x'|\big)\big\}
\end{equation}
for fixed $\alpha\in(0,1]$, with the topology of uniform convergence on compact subsets of $[0,T^\ast]\times\mathbb R$.

\begin{assumption} \label{ass:calib} \NL
\begin{enumerate}
\item[(B1)] The numbers of modes are fixed. \Cref{ass:std} holds uniformly over $(L,\vartheta)\in\mathfrak L\times\mathcal K$. The parameters $a_i$, $\gamma_i$, $\lambda_j$, $q_i$ and $\rho_i$ are continuous in $\vartheta$, with positive lower bounds for $a_i,\gamma_i,\lambda_j$ and a uniform bound $\sum_i\rho_i^2 \le \rho_\ast^2<1$. The functions $U'$, $F$, $c_j$ and
\[
\theta(t,\mathbf z;L,\vartheta) = \left[\mu(t,\mathbf z;L,\vartheta) - (r_t-d_t)\right] / \varsigma(t,\mathbf z; L,\vartheta)
\]
are continuous in $(L,\vartheta)$, uniformly on compact sets of $(t,\mathbf z)$, and locally Lipschitz in the state with constants uniform in $(t,L,\vartheta)$. Functions that are fixed across the family satisfy the corresponding continuity requirement trivially. Writing $G_\vartheta$ for \cref{eq:lyapunov}, assume also that, for constants $c_\ast>0$ and $C_\ast<\infty$, $G_\vartheta(\mathbf z)\ge c_\ast|\mathbf z|^2-C_\ast$ uniformly in $\vartheta$, and $\sup_\vartheta \sup_{\mathbf z\in K} G_\vartheta(\mathbf z)<\infty$ for every compact state set $K$.

\item[(B2)] The premium is $\boldsymbol\Lambda(t,\mathbf z;L,\vartheta)$, bounded uniformly by $\bar\Lambda$, locally Lipschitz in $\mathbf z$ uniformly in $(t,L,\vartheta)$, orthogonal to $\mathbf r$, and continuous in $(L,\vartheta)$ uniformly on compact sets of $(t,\mathbf z)$.

\item[(B3)] The initial state $\mathbf z_0$ is fixed.
\end{enumerate}
\end{assumption}

For a parametric premium, (B2) is a condition on the parametrization. For the utility-selected premium, it asks for the feedback form of \cref{prop:feedback} with a bounded, locally Lipschitz premium for every admissible pair, and for continuity of that premium in $(L,\vartheta)$. \Cref{thm:memm} gives the premium as a BMO process for every pair, but not these uniform Markovian properties, which we assume. The local function enters \cref{eq:bsde} through the Sharpe ratio and the dynamics of the state, so the selected premium can change with $L$ even though the algebraic form \cref{eq:residual_variance} does not.

The theorem below is therefore a theorem about a parametric family of premiums satisfying (B2). For the utility-selected premium, (B2) is checkable only through the additional regularity of \cref{prop:feedback}: the premium has the feedback form when \cref{eq:ptoq_omega} admits a bounded classical solution depending on $(u,\zeta)$ only through $h$, and the uniform bounds in $(L,\vartheta)$ must be verified on the admissible family. When these hold, the theorem applies to the utility-selected premium. When they do not, the theorem applies to a parametric family chosen to satisfy (B2), and the calibration selects within that family. The two readings are distinct, and the theorem should be read with the appropriate one in mind.

\begin{lemma}[Continuity of prices] \label{lem:continuity}
Under \cref{ass:calib}, the maps $(L,\vartheta)\mapsto C_k(L,\vartheta)$ and $(L,\vartheta)\mapsto P_m(L,\vartheta)$ are continuous on $\mathfrak L\times\mathcal K$.
\end{lemma}

\begin{proof}
See \cref{app:lemma}.
\end{proof}

Let $\mathcal R_L:\mathfrak L \to [0,\infty]$ be lower semicontinuous, for example
\begin{equation} \label{eq:local_regularizer}
\mathcal R_L(L) = \int_{\mathcal D}\big[\ell^2  +  b_t|\partial_t\ell|^2  +  b_x|\partial_x\ell|^2  +  b_{xx}|\partial_{xx}\ell|^2\big]\,\mathfrak w(t,x)\,dt\,dx, \qquad \ell = \log\Lh,
\end{equation}
on a bounded Lipschitz domain $\mathcal D$, with fixed nonnegative coefficients and a fixed weight satisfying $0<\mathfrak w_- \le \mathfrak w \le \mathfrak w_ + <\infty$. Only derivatives whose coefficients are strictly positive are required to belong to $L^2(\mathcal D)$, and otherwise the functional is $ + \infty$. A term with zero coefficient is omitted together with its derivative requirement. Let $\mathcal D_{\PP}:\mathcal K \to [0,\infty)$ be continuous, for example a negative log-likelihood or a quadratic distance to physical-measure estimates $\hat\vartheta_{\PP}$.

Consider
\begin{equation} \label{eq:band_problem}
\min\ \mathcal J(L,\vartheta) = w_1\mathcal R_L(L)  +  w_2\mathcal D_{\PP}(\vartheta)  +  w_3\sum_m\frac{(P_m(L,\vartheta) - P_m^{\mathrm{mkt}})^2}{s_m^2}
\end{equation}
over the feasible set
\begin{equation} \label{eq:band}
\mathcal F = \big\{(L,\vartheta)\in\mathfrak L\times\mathcal K:\ C_k^{\mathrm{bid}} \le  C_k(L,\vartheta) \le  C_k^{\mathrm{ask}}\ \text{for all}\ k\big\},
\end{equation}
with weights $w_1,w_2,w_3 > 0$ and scales $s_m > 0$.

\begin{theorem}[Existence of a band calibration] \label{thm:band}
Under \cref{ass:calib}, if $\mathcal F$ is nonempty and $\mathcal J$ is finite at some point of $\mathcal F$, then $\mathcal J$ attains its minimum on $\mathcal F$. The theorem applies to any parametric family of premiums satisfying (B2). When the utility-selected premium satisfies (B2), the theorem applies to it.
\end{theorem}

\begin{proof}
The family \cref{eq:admissible_L} is uniformly bounded and equicontinuous on compact sets, and it is closed under uniform convergence on compact sets. By the Arzel\`a--Ascoli theorem and a diagonal argument, $\mathfrak L$ is compact and metrizable, and so is $\mathfrak L\times\mathcal K$ because $\mathcal K$ is compact. By \cref{lem:continuity}, $\mathcal F$ is closed, hence compact. The functional \cref{eq:local_regularizer} is lower semicontinuous: if $L_n \to  L$ and $\liminf\mathcal R_L(L_n) < \infty$, choose a subsequence attaining this lower limit. The derivatives of $\ell_n$ with strictly positive penalty coefficients are then bounded in weighted $L^2(\mathcal D)$. Along a further common subsequence they converge weakly. Local uniform convergence, the uniform bounds on $L_n$, and testing against compactly supported smooth functions identify their weak limits as the corresponding distributional derivatives of $\ell$. Weak lower semicontinuity of the squared norms, and convergence of the zeroth-order term, give the claim. No unpenalized derivative is required or used. The other two terms of $\mathcal J$ are continuous. A lower semicontinuous function attains its minimum on a nonempty compact set.
\end{proof}

The theorem requires feasibility and does not assert it. The intervals alone do not guarantee that a local function satisfying \cref{eq:band} exists, nor that the minimizer is unique. The problem sidesteps the McKean--Vlasov fixed point, since $L$ is a variable of the optimization rather than a functional of the law. A related variational approach, based on semimartingale optimal transport and finitely many option prices, is developed for local stochastic volatility models in \cite{GuoLoeperWang2022}. The selection among feasible local functions is made by the regularizer, which favors a small and smooth correction, and by $\mathcal D_{\PP}$, which keeps the structural parameters near their physical-measure estimates. Replacing the hard constraints by a continuous penalty on violations of the band gives an unconstrained problem on $\mathfrak L\times\mathcal K$, which has a minimizer by the same argument whenever its objective is finite somewhere. Calibration to a corridor rather than to a point is related to \cite{Avellaneda1997, Guyon2024}.

\begin{corollary}[Vanishing band] \label{cor:vanishing}
Let $\mathcal F_n$ be the feasible sets for bands $[C_k^{\mathrm{mid}} - \epsilon_n, C_k^{\mathrm{mid}}  +  \epsilon_n]$ with $\epsilon_n\downarrow 0$, and assume that the exact set $\mathcal F_0 = \{C_k(L,\vartheta) = C_k^{\mathrm{mid}}\ \text{for all}\ k\}$ is nonempty with $\mathcal J$ finite somewhere on it. Then every sequence of minimizers of $\mathcal J$ on $\mathcal F_n$ has limit points, and each limit point minimizes $\mathcal J$ on $\mathcal F_0$.
\end{corollary}

\begin{proof}
By compactness a subsequence converges to some $(L^\ast,\vartheta^\ast)$, which lies in $\mathcal F_0$ by \cref{lem:continuity}. Since $\mathcal F_0\subset\mathcal F_n$, lower semicontinuity gives $\mathcal J(L^\ast,\vartheta^\ast) \le \liminf_n\min_{\mathcal F_n}\mathcal J \le \inf_{\mathcal F_0}\mathcal J$.
\end{proof}

The band problem is therefore a regularized relaxation of exact calibration to the quoted strikes, and it selects, in the limit, the exact calibration with the smallest and smoothest correction and the structural parameters closest to the physical-measure information.

\subsection{An admissible specification} \label{sec:example}

The assumptions are satisfied by the following specification. Take
\begin{equation} \label{potential}
U(y) = \tfrac12 a_U y^2  +  \tfrac16 c_3 y^3  +  \tfrac1{24} c_4 y^4
\end{equation} with $a_U\in\mathbb R$ and $c_4 > 0$, so that $-yU'(y) \le  -k_1y^2  +  k_2$ and the cubic coefficient $c_3 = U'''(0)$ controls the asymmetry of the driver. Take $f(v) = \sqrt v$ and $\phi(y) = v_{\min}  +  \tfrac12(y  +  \sqrt{y^2  +  \epsilon^2})$, for which $F^4 \le  C(1  +  y^2)$. Take leverage amplitudes \cref{eq:cj_form} with $\psi = \tanh((\cdot - y_\star)/s_\star)$ and $0 \le \eta < 1$. Let $\log\Lh$ belong to a bounded set of a finite-dimensional space of smooth functions with bounded derivatives, and let the physical drift have a constant Sharpe ratio, $\mu = r - d  +  \bar\theta\varsigma$.

Then \cref{ass:std} holds. If, in addition, $t\mapsto r_t-d_t$ is smooth, all coefficients are $C^\infty$, and \cref{prop:hypo} applies when the leverage rates are distinct. A premium from a finite-dimensional family of bounded smooth functions, orthogonal to $\mathbf r$ and depending continuously on its parameters, satisfies (B2) when its local derivative bounds are uniform. Restrict the parameters to compact sets away from $c_4=0$, zero relaxation rates and $|\rho|=1$, and keep the other positive scales bounded away from zero. The polynomial potential then gives uniform coercivity and all the conditions in (B1). The band theorem applies whenever there is a feasible point with finite objective.

For the stated constant-Sharpe-ratio drift, \cref{cor:deterministic_sharpe} gives the utility-selected premium explicitly as $\boldsymbol\Lambda^0=0$. Its regularity and the stability in (B2) are therefore automatic. This example does not produce a nonzero residual memory-risk premium. Without leverage, and with one intrinsic mode or modes uncorrelated with returns, \cref{prop:noleverage} gives a premium in feedback form for a state-dependent Sharpe ratio, and only the uniform properties in (B2) remain to be checked. \Cref{prop:ou_leverage} does the same with constant leverage for its tuned family. With leverage and a general state-dependent Sharpe ratio, \cref{thm:memm} gives the premium, while its feedback form and the uniform properties in (B2) remain additional requirements.

\section{Relation to rough LSV} \label{sec:brigolucic}

The computation of $\EE[F^2(Y_t)|S_t = K]$ is where the GLE--LSV model and rough LSV models differ. In a rough model the variance is non-Markovian, and there is no finite state consisting of the current price and variance. One route conditions on the noise that drives the variance. In a one-factor benchmark, the volatility is adapted to a Brownian motion $W^V$, the price is driven by $\rho W^V  +  \sqrt{1 - \rho^2}W^\perp$ with $W^\perp$ independent, and the conditional density $p_t$ of $S_t$ given $\mathcal F^{W^V}_t$ is defined by $\langle p_t,\varphi\rangle = \EE[\varphi(S_t)|\mathcal F^{W^V}_t]$. With $\Sigma_t(s) = sL(t,s)f(V_t)$ and $\Gamma_t = \rho\Sigma_t$, it solves in weak form
\begin{equation} \label{eq:bl_spde}
d\langle p_t,\varphi\rangle = \Big\langle p_t,\,(r_t - d_t)s\varphi'  +  \tfrac12\Sigma_t^2\varphi''\Big\rangle\,dt  +  \big\langle p_t,\,\Gamma_t\varphi'\big\rangle\,dW^V_t,
\end{equation}
a stochastic partial differential equation whose coefficients depend on the volatility path, \cite{CrisanPardoux2026, BuginiFrizLeZhang2025, DasGuoLoeper2026}. An It\^o--Wentzell transformation along the flow generated by $\Gamma_t$ removes the transport term and leaves, for each fixed path of $W^V$, a Fokker--Planck equation with path-dependent coefficients, \cite{BrigoLucic2026}. Pathwise formulations lead to rough partial differential equations with stochastic coefficients, \cite{FrizHocquetLe2026, BuginiFrizStannat2025, BankBayerFrizPelizzari2025}. The conditional expectation is then assembled from a conditional density.

\begin{myremark}[Conditional representation] \label{rem:raoblackwell}
Let $\mathcal G$ be a sub-$\sigma$-field such that $F^2(Y_t)$ is $\mathcal G$-measurable and the law of $S_t$ given $\mathcal G$ has a density $p_t(\cdot\,|\,\mathcal G)$, and let $\EE[F^2(Y_t)] < \infty$. For bounded Borel $\varphi$, Fubini's theorem gives
\[
\EE[F^2(Y_t) \varphi(S_t)] = \int \varphi(K)\, \EE[F^2(Y_t) p_t(K| \mathcal G)]\,dK,
\]
and the density of $S_t$ is $\EE[p_t(\cdot|\mathcal G)]$. Hence, for almost every $K$ with $\EE[p_t(K|\mathcal G)] > 0$,
\begin{equation} \label{eq:raoblackwell}
\EE\big[F^2(Y_t)\,\big|\,S_t = K\big] = \frac{\EE\big[F^2(Y_t)\,p_t(K|\mathcal G)\big]}{\EE\big[p_t(K|\mathcal G)\big]}.
\end{equation}
The identity rewrites the conditional expectation in \cref{eq:gyongy} as a ratio of two unconditional expectations, both of which involve the same conditional density.

It is the form in which the conditional expectation is computed in a rough or non-Markovian model: one draws the volatility path, computes the conditional density of $S_t$ given that path, and averages over the conditioning field. The Gaussian or Markovian structure of the model enters only through the density $p_t(\cdot\, | \, \mathcal G)$. The identity itself is distribution-free. The right-hand side is well defined whenever the denominator is positive, and it is the natural object to estimate by simulation when a closed form for the conditional density is not available.
\end{myremark}

In the GLE--LSV model, the identity in \cref{rem:raoblackwell} does not apply directly. Since variance responds to returns, the intrinsic noises alone do not determine the path of $Y$, and $F^2(Y_t)$ is therefore not measurable with respect to this choice of $\mathcal G$. A conditioning field that determines $Y_t$ would also need to include the leverage states, but the conditional law of $S_t$ given this larger field is not known to admit a smooth density. The lifted model avoids the need for this identity. Under the hypotheses of \cref{prop:hypo} and its pricing extension in \cref{sec:reduced_Q}, the conditional expectation can be expressed as a ratio of integrals involving a smooth density that solves a standard forward equation, as in \cref{eq:condexp_lift}. Alternatively, it can be represented as the expectation of a coordinate of a finite-dimensional Markov process and evaluated by simulation.

A finite exponential approximation of the fractional kernel, \cite{AbiJaberElEuch2019, BayerBreneis}, also produces a Markov state and an ordinary forward equation, so three cases should be distinguished. In the exact rough model, the conditional expectation passes through \cref{eq:bl_spde} or a rough partial differential equation. A finite lift of the rough model removes this step but approximates a kernel whose short- and long-time power laws share one exponent. At fixed dimension the approximation has neither an actual singularity at zero nor an algebraic tail. A deterministic fractional kernel by itself does not supply the current-state leverage amplitudes studied here, and adding them is a separate modeling choice.

The GLE--LSV model obtains the same ordinary forward equation together with a kernel with independent short-time and long-time regimes, a potential that is a modeling choice, leverage with state-dependent amplitudes, and parameters estimated under the physical measure. The finite-dimensional forward equation is shared with any finite lift. What distinguishes the present construction is that it keeps this formulation together with these controls.

\section{Numerical illustration} \label{sec:numexp}

We illustrate the leverage mechanism with a Monte Carlo experiment on the reduced one-mode system of \cref{prop:reduced}. The purpose is not to calibrate the model, but to show how leverage memory, its state dependence, and the return convention shape the response of the variance driver to a drawdown and to a rally of the index. All runs satisfy the assumptions of \cref{thm:wellposed} or \cref{prop:arithmetic}, so the solutions are global and the experiment does not probe explosion.

\subsection*{Setup}

We take $N=M=1$, $a=\gamma=1$, $\lambda=0.2$ and $\Theta=0.5$, giving the noise loading $q=\sqrt{2\Theta a\gamma}=1$. We set $\rho_1=0$, so returns act on the driver only through the leverage state. The local function is $L\equiv1$, and the short rate and the dividend yield vanish. We use the potential in \eqref{potential} with $a_U=1, c_3=0, c_4=1$, which satisfies \cref{ass:std}(A1) and confines the driver through a cubic restoring force. The variance map is the admissible specification of \cref{sec:example},
\[
\phi(y) = v_{\min}  +  \tfrac12\bigl(y + \sqrt{y^2 + \epsilon^2}\bigr), \qquad v_{\min}=0.04,\quad \epsilon=0.1,
\]
with $F=\sqrt\phi$ and $f(v) = \sqrt v$, therefore \cref{eq:growth_link} holds with a margin. The state-dependent amplitude follows \cref{eq:cj_form},
\[
c(y) = c_0\bigl(1  +  \eta\tanh((y-y_\star)/s_\star)\bigr), \qquad c_0=0.5,\quad \eta=0.9,\quad y_\star=0,\quad s_\star=0.5.
\]
Since $\eta<1$, the amplitude stays in $(0.05,0.95)$ and satisfies (A3), and since $y_\star=Y_0=0$, it equals the constant amplitude $c_0$ at the initial state. The shock is imposed through the physical drift, with $\mu=-2.5$ on the window $[0.4,0.8]$ for a drawdown, $\mu= + 2.5$ on the same window for a rally, and $\mu=0$ elsewhere, so that the log-price has drift $\mu-\tfrac12\varsigma^2$. The initial state is $X_0=0$, $Y_0=0$ and $h_0=g_0=0$. The horizon is $T=2$, the Euler--Maruyama step is $\Delta t=4\times10^{-4}$, that is, $n=5{,}000$ steps, and each ensemble has $5{,}000$ paths.

\begin{figure}[!htb]
\centering
\includegraphics[width=\textwidth]{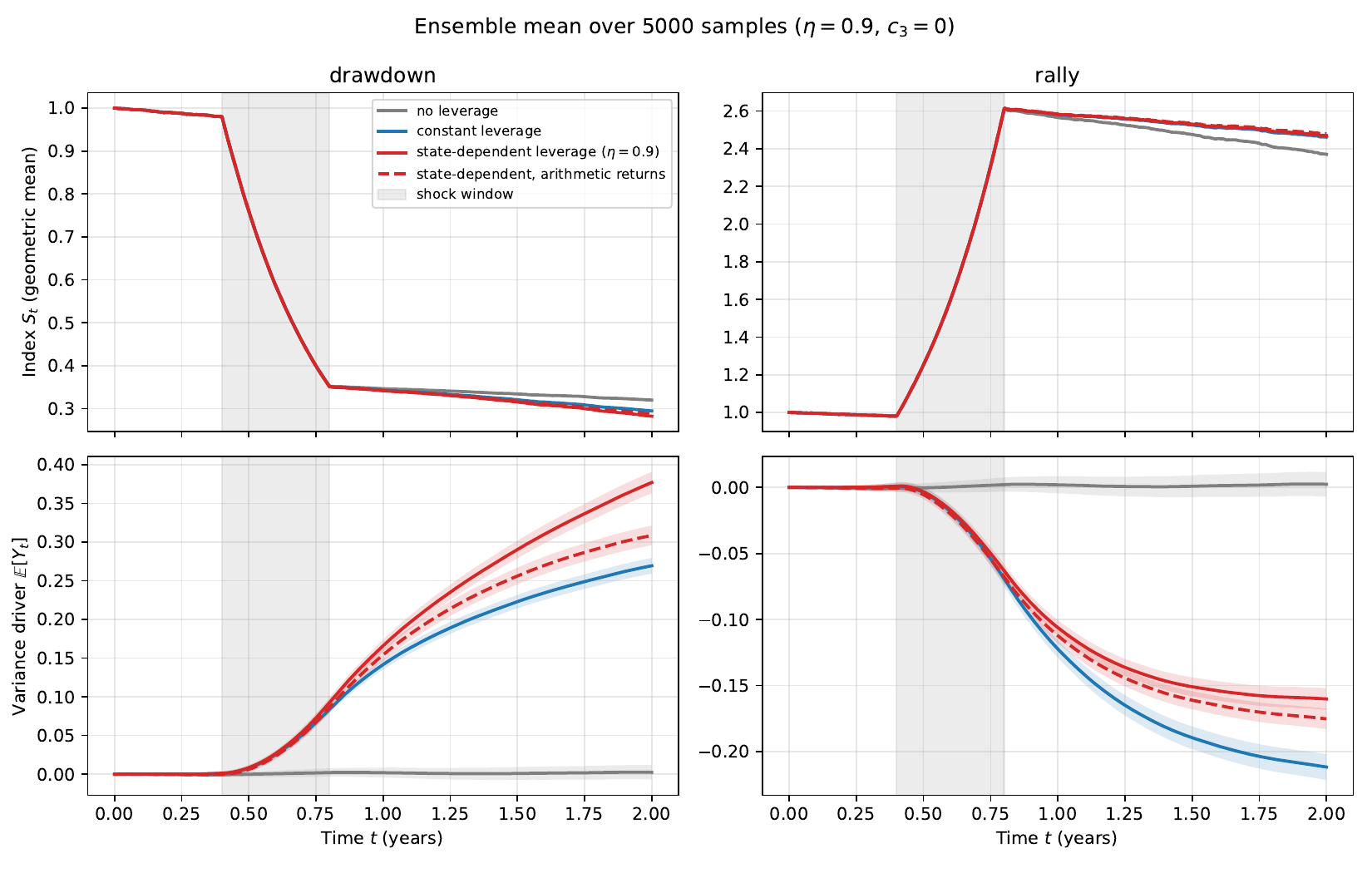}
\caption{Ensemble means of the reduced one-mode GLE--LSV system with $c_3=0$ under a drawdown (left) and a rally (right) of the index, with the shock window shaded. \emph{Top:} geometric mean $\exp(\EE[X_t])$ of the index. \emph{Bottom:} $\EE[Y_t]$ with bands of two standard errors, for no leverage (gray), constant leverage (blue), state-dependent leverage (red), and state-dependent leverage with arithmetic returns (red, dashed). Parameters as in the text.}
\label{fig:numexp}
\end{figure}

Four leverage specifications are compared under common random numbers, so that the Brownian increments are identical across runs and any difference in $Y_t$ is due to the leverage term:
\begin{enumerate}
\item[(i)] \emph{No leverage}, $c\equiv0$.
\item[(ii)] \emph{Constant leverage}, $c\equiv c_0$.
\item[(iii)] \emph{State-dependent leverage}, $c(y)$ as above.
\item[(iv)] \emph{State-dependent leverage with arithmetic returns}, $c(y)$ as above, with the leverage state driven by $dS_t/S_t$ instead of $dX_t$.
\end{enumerate}
The top panels of \cref{fig:numexp} report the geometric mean $\exp(\EE[X_t])$ of the index. The arithmetic mean $\EE[S_t]$ does not distinguish the specifications. The drift $\mu$ is deterministic, and $S_t\exp(-\int_0^t\mu_s\,ds)$ is a true martingale by the argument of \cref{lem:martingale}, so $\EE[S_t]=S_0\exp(\int_0^t\mu_s\,ds)$ in every case. In the runs, $\EE[S_T]$ is $0.370$ after the drawdown in all four specifications. The bottom panels report $\EE[Y_t]$ with bands of two standard errors.

\subsection*{Interpretation}

Without leverage the driver does not see the index, and $\EE[Y_t]$ stays at zero up to Monte Carlo error. With leverage, the leverage state $g_t$ turns negative during the drawdown, and the term $-c(Y_t)g_t$ raises the drift of $Y$. Because $\lambda=0.2$, the part of the drawdown accumulated over the window that remains in the leverage state at the horizon is $\tfrac{1}{0.4}\int_{0.4}^{0.8}e^{-0.2(2-s)}\,ds\approx0.76$, and the driver relaxes slowly, so $\EE[Y_t]$ is still rising at $T=2$. With constant leverage $\EE[Y_T]=0.269$ after the drawdown, and a rally of the same size gives $\EE[Y_T]=-0.212$.

State dependence breaks the symmetry between the two responses. Since $y_\star=Y_0$, the state-dependent amplitude exceeds $c_0$ once the driver has risen and falls below $c_0$ once it has fallen. The drawdown response therefore grows to $\EE[Y_T]=0.377$, about $40\%$ above the constant case, and the rally response shrinks to $\EE[Y_T]=-0.160$, about $25\%$ below it. Without a shock the driver does not stay at zero, because the It\^o drift of the log-price enters the leverage state, and $\EE[Y_T]$ equals $0.027$ for constant and $0.046$ for state-dependent leverage. Let $m_-$, $m_+$ and $m_0$ denote $\EE[Y_T]$ after a drawdown, after a rally and without a shock. The centered contrast $m_- + m_+-2m_0$ vanishes for a response that is odd in the shock. It equals $0.003$ for constant leverage and $0.124$ for state-dependent leverage. The standard errors of the means are at most $0.007$, and those of the contrasts, computed path by path under common random numbers, are at most $0.001$. With constant leverage the response is thus essentially odd, and the asymmetry comes from the state dependence of the amplitude. This is the sense in which a positive $\eta$ strengthens the leverage in stress and weakens it in calm.

The return convention accounts for part of the response. With arithmetic returns the leverage state omits the It\^o drift $-\tfrac12\varsigma^2$ of the log-price, and the state-dependent drawdown response falls from $0.377$ to $0.309$, while the rally response becomes $-0.175$. The no-shock value falls from $0.046$ to $0.012$, and the centered contrast from $0.124$ to $0.109$. With log returns this drift enters the leverage state after both shocks and pushes the driver up, which is the mechanism described in \cref{rem:growth}. In the present global regime the effect is moderate. It becomes destabilizing only for the faster growth of $F$ identified in \cref{thm:dichotomy}.

The index panels differ less than the driver panels, because the leverage term enters the drift of $Y$ and reaches the index only through $\varsigma(Y)$. The difference becomes visible after the shock, since the geometric mean of the index drifts at the rate $-\tfrac12\EE[\varsigma_t^2]$ and a higher driver lowers it. At $T=2$ the drawdown leaves $\exp(\EE[X_T])$ at $0.320$ without leverage and at $0.282$ with state-dependent leverage. A joint SPX--VIX calibration, rather than an SPX-only calibration, is therefore the natural setting in which to identify the leverage amplitudes. The corresponding estimation problem is taken up in the companion paper.

\subsection{Asymmetric potential}

\begin{figure}[!htb]
\centering
\includegraphics[width=0.8\textwidth]{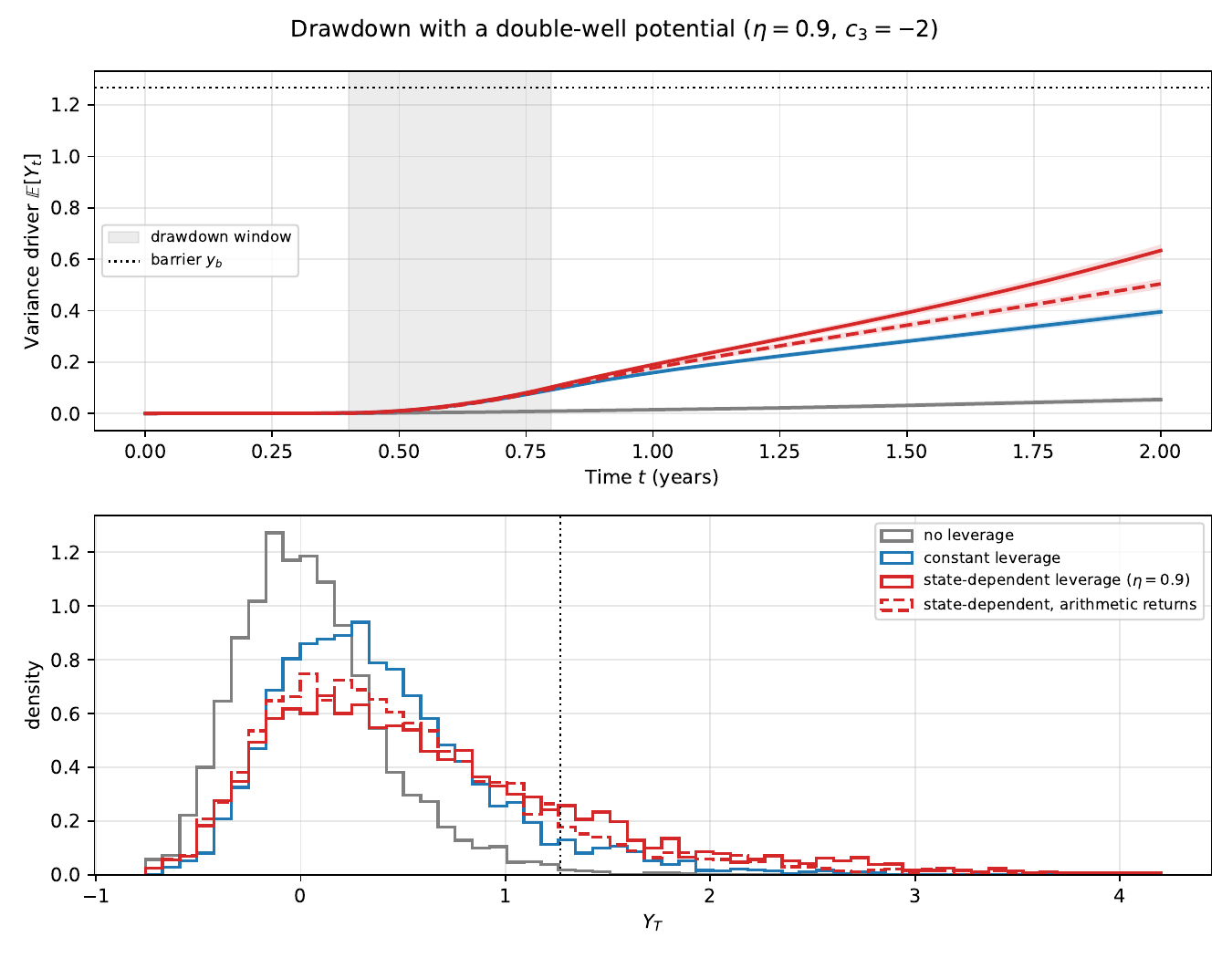}
\caption{Drawdown experiment of \cref{fig:numexp} with $c_3=-2$. \emph{Top:} $\EE[Y_t]$ with bands of two standard errors, and the barrier $y_b\approx1.27$ (dotted). \emph{Bottom:} histograms of $Y_T$ at $T=2$. Colors as in \cref{fig:numexp}.}
\label{fig:numexp_c3}
\end{figure}

The experiment above sets $c_3=0$, so that the potential is symmetric about the origin. The admissible specification of \cref{sec:example} allows a nonzero cubic coefficient, which introduces an independent source of asymmetry. \Cref{fig:numexp_c3} repeats the drawdown experiment with $c_3=-2$ and the other parameters unchanged. This potential satisfies (A1), and since $c_4>0$, it also satisfies the lower bound on $U''$ in \cref{thm:dichotomy}(i). It has two local minima, at $y=0$ and at $y=3 + \sqrt3\approx4.73$, separated by a local maximum at $y_b=3-\sqrt3\approx1.27$ with barrier height $U(y_b)-U(0)\approx0.23$. The right well is much deeper, with $U(3 + \sqrt3)\approx-3.23$.

Without leverage only $0.6\%$ of the paths satisfy $\sup_{t \le  T}Y_t>y_b$, so on this horizon the noise alone rarely moves the driver between the wells, and the transition is driven by the leverage term. After the drawdown, the fraction of paths that cross the barrier at least once is $6.6\%$ for constant leverage, $18.6\%$ for state-dependent leverage, and $12.3\%$ for state-dependent leverage with arithmetic returns. The fractions that end in the right well, $Y_T>y_b$, are almost the same, $6.6\%$, $18.5\%$ and $12.1\%$, so few paths that cross return by $T$. State dependence thus nearly triples the crossing probability, and the log-return convention raises it from $12.3\%$ to $18.6\%$. After a rally of the same size, at most $0.3\%$ of the paths cross in every specification. Even without a shock, leverage raises the crossing fraction to $1.5\%$, $3.7\%$ and $1.8\%$. Halving the time step on the same Brownian paths changes these fractions and the means by less than $0.001$. The means $\EE[Y_T]=0.395$, $0.634$ and $0.504$ in the three leveraged cases mix the paths that remain in the left well with those that end in the right well, as the histograms in \cref{fig:numexp_c3} show.

The cubic term does not change the growth condition of \cref{thm:dichotomy}, which depends only on the growth of $U$ and $F$ at infinity. Its effect in the experiment is the transition between the wells, which the leverage triggers and its state dependence amplifies. Whether this mechanism can be separated from a stronger leverage in a symmetric potential is part of the estimation problem of the companion paper.

\section{Conclusion} \label{sec:conclusion}

The GLE--LSV construction distinguishes intrinsic volatility memory from the influence of past returns and allows the latter to vary with the variance state. The main analytical difficulty is that the It\^o correction in log returns can reinforce the variance driver through leverage memory. We establish well-posedness under a growth bound linking the volatility map to the potential, and we show that some such bound is necessary. In a one-mode family, explosion is possible from suitable initial states exactly when the squared volatility factor outgrows the restoring force, and otherwise the solution is global for every initial state. Arithmetic returns relax the condition for existence but not the condition used for the true martingale property.

For finite exponential kernels, we reduce the state by combining matched friction and noise modes and verify a bracket condition yielding a smooth density. Under a bounded Sharpe ratio, the measure selected by exponential utility is the minimal entropy martingale measure, and its premium is obtained from a quadratic BSDE. When the premium has a smooth feedback form, the reduced representation is preserved under the selected measure. For bounded Markovian premiums, the discounted total-return index is a true martingale, and SPX and VIX instruments share a common linear valuation rule. The growth estimates, the explosion criterion and the bracket calculation describe how return feedback affects the probabilistic foundation of the model. The pricing and inverse-problem results use that foundation under their additional hypotheses.

The finite-kernel restriction is an important boundary of the present results. The structural kernels have a singular origin and algebraic tails, whereas each finite representation has a regular origin and exponential decay. Uniform kernel approximation on a compact interval of positive lags does not by itself establish convergence of the associated stochastic systems, pricing measures, or derivative prices. Extending the theory as the number of modes increases requires estimates uniform in the approximation and a suitable interpretation of the limiting colored force. Questions about feedback also remain. In the one-mode family the growth threshold for log returns is sharp and lies above the range covered by the general growth condition. Whether that condition can be weakened with several modes or with state-dependent amplitudes, and what happens for arithmetic returns beyond the growth of the potential, remain open.

The utility-based construction yields the selected measure from primitive assumptions, with the premium as a BMO process. What remains is its regularity. A bounded Markovian feedback form, which the true martingale property and the reduced representation under the selected measure use, requires a classical solution of the semilinear equation with bounded exposures. The inverse-problem theorem also assumes continuity of the selected premium in the local function and the structural parameters. Stability results for quadratic BSDE can give convergence of the premium processes in integral norms, but not the uniform local bounds assumed there. The H\"ormander condition itself is proved under the stated smoothness and nondegeneracy hypotheses. It does not, however, provide the global estimates required for the nonlinear pricing equation.

State-dependent intrinsic memory provides a different extension. Matched multiplicative noise and friction, or suitable conditional covariance relations, would require a revised specification and new estimates. The restriction noted here concerns only the pathwise identity equating an unconditional covariance with a random exponential kernel, and it does not exclude those broader models. Without leverage, and with a single unhedgeable direction or modes uncorrelated with returns, a distortion transform gives such a solution for the potentials covered by \cref{prop:noleverage}. With constant leverage, \cref{prop:ou_leverage} gives one in a tuned family where a combination of the state variables is an autonomous Ornstein--Uhlenbeck factor. For general Sharpe ratios with leverage, whether a bounded classical solution of the semilinear equation exists under primitive assumptions on the coefficients is an open question, which the paper treats as a hypothesis.

For the inverse problem, existence of a minimizer is only the first question. Feasibility of the quote constraints, uniqueness or identification of the structural parameters, and quantitative stability under perturbations of prices remain unresolved. In particular, smoothness of the joint density does not provide a uniform lower bound for the spot density in the wings, where conditional-expectation formulas can be sensitive. The vanishing-band result concerns a fixed finite collection of quotes and does not establish solvability of exact calibration to an entire surface. Long-time behavior is another separate problem: the present finite-horizon estimates do not establish invariant distributions or ergodicity. Since the local function depends on the nonstationary price, any stationary formulation for volatility and memory must specify the additional restrictions under which that subsystem can be treated autonomously.

The companion paper will address statistical estimation, numerical solution, and calibration to SPX and VIX data. It will ask whether state-dependent leverage can be distinguished from changes in the potential, intrinsic memory, and local volatility function, using historical observations together with derivative prices. Comparisons with constant leverage, with $L\equiv 1$, and with finite approximations of rough models will help evaluate the contribution of each modeling choice. Sensitivity to the number of memory modes and to the assumed noise--friction relation will also be examined. The results proved here supply the probabilistic and pricing foundation for those investigations, while their empirical adequacy and computational performance remain to be established.

\section*{Disclosure statement}

No potential conflict of interest was reported by the author.

\section*{Funding}

No funding was received.

\section*{Disclaimer}

Opinions expressed here are the author's own and do not represent the views of his employer. A standard disclaimer applies.

\section*{Acknowledgments}

I thank Peter Friz and Rakhymzhan Kazbek for helpful discussions and comments.

\printbibliography[title={References}]

\appendix
\label{app:proofs}
\appendixpage
\numberwithin{equation}{section}
\setcounter{equation}{0}

\section{Proof of \cref{thm:wellposed}} \label{app:theorem2}

\begin{proof} \NL
\myparagraph{Local solution.}
The coefficients of \cref{eq:lift_P} are Borel in $t$ and locally Lipschitz in $\mathbf z$ uniformly in $t$. Indeed, $U'$ is $C^1$, the $c_j$ are locally Lipschitz, $\varsigma = \Lh F$ is a product of a bounded function that is Lipschitz in $x$ and a locally Lipschitz function of $y$, and $\varsigma^2$ and $\mu$ inherit local Lipschitz continuity. Hence there is a unique strong solution up to the explosion time $\tau_\infty = \lim_n\tau_n$, with $\tau_n = \inf\{t:|\Zs_t|\ge n\}$, \cite{Protter2005}.

\myparagraph{Lyapunov bound.}
Let $\mathcal L^{\PP}$ denote the generator of \cref{eq:lift_P}. Put $A = U'(y)  +  \sum_iu_i$ and $R = \sum_i\zeta_i - \sum_jc_j(y)g_j$, so that $m = -A  +  R$. Since $G$ is a sum of functions of single coordinates, only diagonal second-order terms contribute, and
\begin{align*}
\mathcal L^{\PP}G &= Am - \sum_i\frac{\gamma_i}{a_i}u_i^2  +  \sum_i\big(q_i^2 - 2\gamma_i\zeta_i^2\big)  +  \sum_j\big[2g_j(-\lambda_jg_j  +  \mu - \tfrac12\varsigma^2)  +  \varsigma^2\big]  +  2x(\mu - \tfrac12\varsigma^2)  +  \varsigma^2.
\end{align*}

First,
\[
Am = -A^2  +  AR \le \tfrac12R^2 \le \tfrac12(N  +  M)\big(\sum_i\zeta_i^2  +  c_{\max}^2\sum_jg_j^2\big).
\]
Next, $2ab \le  a^2  +  b^2$ gives
\[
2g_j(\mu - \tfrac12 \varsigma^2)  \le  g_j^2  +  2\mu^2  +  \tfrac12\varsigma^4,
\]
and likewise for the $x$ term, while (A5) gives $\mu^2 \le  C(1  +  \varsigma^2)$. By (A2) and (A4)
\[
\varsigma^4 \le \ell_{\max}^4C_F(1  +  U - U_\ast), \qquad \varsigma^2 \le \ell_{\max}^2(1  +  \varsigma^4/\ell_{\max}^4).
\]
Collecting terms, $\mathcal L^{\PP}G \le  C_GG$. It\^o's formula and Gronwall's inequality give $\EE^{\PP} [G(\Zs_{t \wedge \tau_n})] \le \EE^{\PP}[G(\Zs_0)]e^{C_Gt}$.

\myparagraph{Non-explosion.}
By (A1), $G(\mathbf z)\ge c|\mathbf z|^2 - C$ for constants $c > 0$ and $C$, so $\inf_{|\mathbf z|\ge n}G \to \infty$. Since $\PP(\tau_n \le  t)\inf_{|\mathbf z|\ge n}G \le \EE^{\PP}[G(\Zs_{t\wedge\tau_n})]$, we obtain $\PP(\tau_\infty \le  t) = 0$ for every $t$, as in Khasminskii's test, \cite{Khasminskii2012}. Fatou's lemma gives \cref{eq:moment_bound}.

\myparagraph{Markov property.} Pathwise uniqueness implies uniqueness in law. The
martingale problem for the space--time process $(t,\Zs_t)$ is therefore well posed, and its solution is strong Markov, \cite{EthierKurtz1986}. Positivity of $S$ and $V$ is immediate.
\end{proof}

\section{Proof of \cref{thm:dichotomy}} \label{app:theorem3}

\begin{proof}
Throughout, $\Phi = F^2$. Since $F$ is nondecreasing and $\inf F>0$, the function $\Phi$ is nondecreasing and bounded below, hence bounded by $\Phi(0)$ on $(-\infty,0]$. All coefficients are locally Lipschitz, so a unique solution exists up to its explosion time $\tau_\infty$. Write
\[
h=\zeta-u, \qquad q=\sqrt{2\Theta a\lambda}, \qquad
\widetilde U(y) = U(y)  +  \tfrac12(\lambda  +  a) y^2, \qquad w=h-cg + (\lambda + a)Y,
\]
so that $\dot Y=w-\widetilde U'(Y)$. Subtraction of the mode equations in \cref{eq:lift_P} gives, up to $\tau_\infty$,
\begin{alignat}{2} \label{eq:dich_pair}
dY_t &= \big(w_t - \widetilde U'(Y_t) \big)\,dt, &\qquad dw_t &= \big( \tfrac c2 \Phi(Y_t) - \lambda U'(Y_t) \big)\,dt  +  dM_t, \\
M_t &= q B_{1,t} - c \int_0^t F(Y_s)\,dB_{0,s}, &\qquad \langle M \rangle_t &= q^2 t +  c^2 \int_0^t\Phi(Y_s)\,ds. \nonumber
\end{alignat}
The pair $(Y,w)$ is thus a closed system. With arithmetic returns the term $\tfrac c2\Phi$ is absent from the drift of $w$, because the drift $-\tfrac12\Phi$ of $X$ no longer enters $g$.

\myparagraph{Proof of (i).}
Assume $p \le  m/2$. On $[1,\infty)$ we have $\Phi(y)^2 \le  C(1  +  y^{2p}) \le  C(1  +  y^m)$, using $2p \le  m$ and $y\ge 1$. On $(-\infty,1]$ the function $\Phi^2$ is bounded. Combining the two, $\Phi^2 \le  C'(1  +  U(y) - U_\ast)$ on all $\mathbb R$, by the lower bound $ky^m \le  U(y) - U_\ast  +  K$. This is \cref{eq:growth_link}. The remaining requirements of \cref{ass:std} hold by the hypotheses on $U$, $F$, and the constant coefficients. Hence \cref{thm:wellposed} applies and the solution is global for every initial state satisfying (A6).

Assume now that $p \le  m-1$ and that $U''$ satisfies the lower bound in (i). Consider
\[
H(y,w)=1 + \lambda\big(\widetilde U(y)-\inf\widetilde U\big) + \tfrac12\bar w^2, \qquad \bar w=w-\widetilde U'(y),
\]
so that $H\ge1$. Since $Y$ has paths of finite variation and $U\in C^2$, It\^o's formula applies to $H(Y,w)$, and by \cref{eq:dich_pair} the drift of $H(Y_t,w_t)$ is $(\mathcal L H)(Y_t,w_t)$ with
\[
\mathcal L H=\lambda(\lambda + a)\,y\,\bar w-\widetilde U''(y)\,\bar w^2 + \tfrac c2\Phi(y)\,\bar w + \tfrac12q^2 + \tfrac12c^2\Phi(y).
\]
For $y\ge\bar y$ we have $\widetilde U''(y)\ge ky^{m-2}>0$, and $2ab \le  a^2 + b^2$ gives
\[
\tfrac c2\Phi\,\bar w + \lambda(\lambda + a)\,y\,\bar w \le \tfrac12\widetilde U''\bar w^2 + \frac{c^2\Phi^2 + 4\lambda^2(\lambda + a)^2y^2}{4\widetilde U''}.
\]
Since $\Phi(y) \le  C_2y^p$, the last term is at most $C(y^{2p-m + 2} + y^{4-m}) \le  C(1 + y^m)$, because $p \le  m-1$ and $m\ge2$. For $y \le \bar y$ the function $\Phi$ is bounded and $\widetilde U''\ge-K$, so
\[
\tfrac c2\Phi\,\bar w  +  \lambda(\lambda  +  a)\,y\, \bar w - \widetilde U''\bar w^2 \le  C(1 + y^2  +  \bar w^2).
\]
Moreover, $\tfrac12c^2\Phi \le  C(1 + y^m)$ on $[1,\infty)$ and is bounded on $(-\infty,1]$. Since $H\ge1$ and $\widetilde U\ge U_\ast + \tfrac12(\lambda + a)y^2$, we have $y^2 + \bar w^2 \le CH$, and the lower bound $ky^m \le U(y)-U_\ast + K$ gives $y^m \le CH$ for $y\ge1$. Collecting terms, $\mathcal LH \le CH$.

As $\widetilde U'$ is continuous, $H(y,w) \to \infty$ as $|(y,w)| \to \infty$. It\^o's formula, Gronwall's inequality and Khasminskii's test, as in the proof of \cref{thm:wellposed}, show that $(Y,w)$ does not explode from any deterministic initial state, and hence from any $\mathcal F_0$-measurable one, because the initial state is independent of the Brownian motions. Given a finite path of $Y$, the coordinates $X$, $g$ and $h=w + cg-(\lambda + a)Y$ are finite on finite intervals, so the solution is global.

\myparagraph{Proof of (ii).}
Fix $T>0$ and $R>0$, and put $\delta=1/(4c)$. The exponential local martingale $\exp(-2\delta M_t-2\delta^2\langle M\rangle_t)$, stopped before explosion, is a nonnegative supermartingale. Applying its maximal inequality at successive exit times and passing to the limit shows that the event
\[
E_R = \{M_t\ge - R - \delta \langle M \rangle_t, \quad \forall\,\, 0 \le  t<T \wedge \tau_\infty \}
\]
has probability at least $1-e^{-2\delta R}>0$. On this event,
\begin{equation} \label{eq:dich_lower}
w_t\ge v_0 + \int_0^t\big(\tfrac c4\Phi(Y_s)-\lambda U'(Y_s)\big)\,ds, \qquad v_0=w_0-R-\delta q^2T.
\end{equation}
Since $p>m-1$, there is $y_1\ge 1$ with $\tfrac c8\Phi(y)\ge\lambda U'(y)$ for $y\ge y_1$. With $D=c/8$, the integrand in \cref{eq:dich_lower} is then at least $D\Phi(Y_s)$ whenever $Y_s\ge y_1$.

Choose $\bar p$ with $1<\bar p<\min\{p-m + 2,(p + 1)/2\}$, which is possible because $p>m-1\ge 1$. Compare with the deterministic system $\dot y=v-\widetilde U'(y)$, $\dot v=D\Phi(y)$. On the curve $v=\widetilde U'(y) + y^{\bar p}$ we have $\dot y=y^{\bar p}$ and, for $y\ge 1$,
\[
\frac{d}{dt}\big(v-\widetilde U'(y)-y^{\bar p}\big)=D\Phi(y)-\big(\widetilde U''(y) + \bar py^{\bar p-1}\big)y^{\bar p}\ge DC_1y^p-(K + \lambda + a)y^{\bar p + m-2}-\bar py^{2\bar p-1}.
\]
Both powers on the right are below $p$, so the right side is positive for $y\ge y_0$ once $y_0\ge y_1$ is large. Hence the region $\{y\ge y_0,\ v\ge\widetilde U'(y) + y^{\bar p}\}$ is forward invariant, and in it $\dot y\ge y^{\bar p}$. Choose $y_0$ also with $y_0^{1-\bar p}/(\bar p-1)<T$. The solution started from $(y_0,v_0)$ with $v_0\ge\widetilde U'(y_0) + y_0^{\bar p}$ then reaches infinity before $T$.

On $E_R$, as long as $Y\ge y_1$, \cref{eq:dich_lower} gives $\dot Y\ge v_\ast-\widetilde U'(Y)$ with
\[
v_\ast(t) = v_0 + D \int_0^t \Phi(Y_s)\,ds, \qquad \dot v_\ast = D \Phi(Y).
\]
The comparison system is cooperative on $y\ge y_1$, since its off-diagonal derivatives are $1$ and $D\Phi'(y)\ge0$. The first-contact comparison argument, localized on bounded rectangles, gives $Y\ge y$ and $v_\ast\ge v$ up to either explosion time. If the stochastic solution survived until $T$, it would remain finite on $[0,T]$ while dominating $y$, which explodes earlier. Hence $\PP(\tau_\infty<T)\ge1-e^{-2\delta R}=1-e^{-R/(2c)}$. The required $w_0$ is obtained by choosing $h_0$ and $g_0$, and then any finite $u_0,\zeta_0$ with $\zeta_0 - u_0 = h_0$.

\myparagraph{Proof of (iii).}
Assume $p \le  m$. On $[1,\infty)$ we have $\Phi(y) \le  C(1  +  y^p) \le  C(1  +  y^m)$, and $\Phi$ is bounded on $(-\infty,1]$. Hence $\Phi \le  C'(1  +  U(y) - U_\ast)$ on $\mathbb R$, and \cref{prop:arithmetic} applies. The solution is global for every initial state satisfying (A6).

\myparagraph{Proof of (iv).}
Assume $c=0$. Then the leverage term vanishes from \cref{eq:gleY_lev}, and the subsystem $(Y,u,\zeta)$ does not involve $F$. Its generator acts on the Lyapunov function
\[
G_0(y,u,\zeta)=1  +  U(y) - U_\ast  +  \sum_i u_i^2/(2a_i)  +  \sum_i\zeta_i^2
\]
through
\[
\mathcal L_0 G_0 = -\Big(U'(y)  +  \sum_i u_i\Big)^2  +  \Big(U'(y)  +  \sum_i u_i\Big)\sum_i\zeta_i - \sum_i\frac{\gamma_i}{a_i}u_i^2  +  \sum_i\big(q_i^2 - 2\gamma_i\zeta_i^2\big),
\]
and the same estimates as in the proof of \cref{thm:wellposed} give $\mathcal L_0 G_0 \le  C_0 G_0$ for a constant $C_0$ depending only on $N,\Theta,\gamma_i,q_i$. No growth condition on $F$ is used. It\^o's formula and Gronwall's inequality bound $\EE[G_0(Y_{t\wedge\tau_n},u_{t\wedge\tau_n},\zeta_{t\wedge\tau_n})]$, and Khasminskii's test gives non-explosion of $(Y,u,\zeta)$. On every finite interval, $S=e^X$ is a stochastic exponential whose coefficients are bounded along the path ($L\equiv1$, $\mu=0$, $F$ locally bounded), so $X$ remains finite, and $g_j$ is an integral of $dX$ against a bounded deterministic kernel. Both are finite on $[0,\tau_\infty)$, so no additional explosion mechanism is present. The argument is identical for the log-return and arithmetic-return conventions, since $c=0$ removes the only term that distinguishes them.
\end{proof}

\section{Proof of \cref{prop:hypo}} \label{app:theorem4}

\begin{proof}
Denote $\mathbf D = \partial_x  +  \sum_j\partial_{g_j}$. In the independent basis, the Stratonovich diffusion fields of \cref{eq:reduced_lift} are
\[
\sigma_k = q_k \partial_{h_k}  +  \rho_k \varsigma\mathbf D,\quad k = 1,\ldots,N, \qquad  \sigma_0 = \sqrt{1 - \rho^2}\, \varsigma \mathbf D.
\]
The It\^o-to-Stratonovich correction $-\tfrac12\sum_k(\sigma_k\cdot\nabla)\sigma_k$ is proportional to $\mathbf D$, because $\varsigma$ does not depend on $h$. Denote the Stratonovich drift by $b$, and put $B=\partial_t + b$, the drift of the space--time process $(t,\Zs^{\mathrm r}_t)$, since the coefficients depend on time. For a smooth function $f$, $[fX,Z] = f[X,Z] - (Zf)X$. Hence the pointwise span of the diffusion fields and of their iterated brackets with one another and with $B$, which is the span in the parabolic H\"ormander condition, does not change when the diffusion fields are multiplied by smooth nonvanishing functions. We may therefore use $\partial_{h_k} = q_k^{-1}(\sigma_k - \rho_k(1 - \rho^2)^{-1/2}\sigma_0)$ and $\mathbf D = \sigma_0/(\sqrt{1 - \rho^2}\,\varsigma)$.

The $y$-component of $b$ is $\tilde m$, and $\partial_{h_k}\tilde m = 1$. Hence $[\partial_{h_k},B]$ equals $\partial_y$ modulo the span of the $\partial_{h_i}$ and $\mathbf D$. It remains to span the $(x,g)$-coordinates. The log-price and each $g_j$ have the same diffusion coefficients and therefore receive the same It\^o-to-Stratonovich correction. If $b_x$ denotes the $x$-component of $b$, the $g_j$-component of $b$ is therefore $b_x - \lambda_jg_j$. Put $W_1 = [\mathbf D,B]$ and $W_{l + 1} = [W_l,B]$. These fields have no $\partial_t$ component, because $\mathbf D$ has none and the $\partial_t$ component of $B$ is constant. Write $W_l^x$ and $W_l^{g_j}$ for the coefficients of $\partial_x$ and $\partial_{g_j}$ in $W_l$. We show by induction that
\[
W_l^x = \alpha_l, \qquad W_l^{g_j} = \alpha_l  +  p_l(\lambda_j), \qquad p_l(\lambda) = \sum_{k=1}^{l}c_{l,k}\lambda^k,
\]
for smooth functions $\alpha_l$ and $c_{l,k}$ that do not depend on $j$, with $c_{l,l} = (-1)^l$.

Since $\mathbf D$ has constant coefficients, each coefficient of $[\mathbf D,B]$ is the derivative along $\mathbf D$ of the corresponding component of $b$. Thus
\[
W_1^x = \mathbf D b_x, \qquad W_1^{g_j} = \mathbf D(b_x - \lambda_jg_j) = \mathbf D b_x - \lambda_j,
\]
which is the claim for $l = 1$ with $\alpha_1 = \mathbf D b_x$ and $p_1(\lambda) = -\lambda$. For the induction step, each coefficient of $[W_l,B]$ is $W_l$ applied to the corresponding component of $b$, minus $B$ applied to the corresponding coefficient of $W_l$. Since $W_lg_j = W_l^{g_j}$,
\[
W_{l + 1}^x = W_lb_x - B\alpha_l, \qquad W_{l + 1}^{g_j} = W_lb_x - \lambda_jW_l^{g_j} - B W_l^{g_j} = W_{l + 1}^x - \lambda_j\alpha_l - \lambda_jp_l(\lambda_j) - B p_l(\lambda_j).
\]
Hence
\[
\alpha_{l + 1} = W_lb_x - B\alpha_l, \qquad p_{l + 1}(\lambda) = -\lambda\alpha_l - \lambda p_l(\lambda) - \sum_{k=1}^{l}(B c_{l,k})\lambda^k.
\]
This polynomial has no constant term, and its coefficients do not depend on $j$. Its leading coefficient is $-c_{l,l} = (-1)^{l + 1}$, because $c_{l,l}$ is constant and $B c_{l,l} = 0$. This completes the induction.

Modulo $\mathbf D$, $\partial_y$ and the $\partial_{h_i}$, the field $W_l$ therefore equals $\sum_jp_l(\lambda_j)\partial_{g_j}$. The matrix $(p_l(\lambda_j))_{l,j=1}^M$ is a lower triangular transformation with nonzero diagonal of $(\lambda_j^k)_{k,j=1}^M$, whose determinant is $\prod_j \lambda_j \prod_{j < k}(\lambda_k - \lambda_j)\ne 0$. Together with $\mathbf D$, the fields $W_1,\ldots,W_M$ therefore span the $(x,g)$-coordinates modulo $\partial_y$ and the $\partial_{h_i}$. All fields used are diffusion fields or iterated brackets of diffusion fields with $B$, so the parabolic H\"ormander condition holds. H\"ormander's theorem, \cite{Hormander1967}, applied to the forward operator, whose adjoint has the same principal fields, gives smoothness of the law, which is a distributional solution of the forward equation.
\end{proof}

\section{Proof of \cref{thm:memm}} \label{app:theorem5}

\begin{proof} \NL
(a) Write $\mathfrak g(t,\mathbf v)$ for the integrand of the $ds$-integral in \cref{eq:bsde}. By (A5), $|\theta| \le \bar\theta$, so $\mathfrak g$ is continuous in $\mathbf v$, $|\mathfrak g(t,\mathbf v)| \le  C(1  +  |\mathbf v|^2)$, and
\[
|\mathfrak g(t,\mathbf v) - \mathfrak g(t,\mathbf v')| \le  C(1  +  |\mathbf v|  +  |\mathbf v'|)|\mathbf v - \mathbf v'|.
\]
The filtration is Brownian and the terminal value is bounded, so \cite{Kobylanski2000} gives a bounded solution, unique among bounded solutions, together with a comparison theorem. For bounded solutions $\int\mathbf V\cdot d\mathbf B$ is BMO, \cite{HuImkellerMuller2005}. The driver lies between $-\theta\,\mathbf v\cdot\mathbf r - \tfrac12\theta^2$ and $\tfrac12|\mathbf v^\perp|^2 - \theta\,\mathbf v\cdot\mathbf r$. The equation with the upper driver has the solution $(0,0)$. The equation with the lower driver is linear, and Girsanov's theorem with the bounded drift $-\theta\mathbf r$ gives its solution
\[
\EE^{\QQ_0} \left[ -\tfrac12\int_t^{T_H}\theta_s^2\,ds\,|\,\mathcal F_t \right] \ge - \tfrac12 \bar\theta^2(T_H - t).
\]
The comparison theorem gives the bounds.

(b) Since $|\boldsymbol\Lambda^0| \le |\mathbf V|$ and $|\theta| \le \bar\theta$, the martingale $N = \int(-\theta\mathbf r  +  \boldsymbol\Lambda^0)\cdot d\mathbf B$ is BMO, so $\mathcal E(N)$ is a positive uniformly integrable martingale by Kazamaki's criterion, \cite{Kazamaki1994}. As $\mathbf r\cdot(-\theta\mathbf r  +  \boldsymbol\Lambda^0) = -\theta$, the process $\tilde S$ is a $\QQ^{(0)}$-local martingale, as in \cref{prop:emm}. Under $\QQ^{(0)}$, $N - \langle N\rangle$ is again BMO, \cite{Kazamaki1994}, so
\[
\EE^{\QQ^{(0)}} [\langle N \rangle_{T_H}] < \infty, \qquad H(\QQ^{(0)}|\PP) = \EE^{\QQ^{(0)}} [N_{T_H} - \tfrac12\langle N \rangle_{T_H}] = \tfrac12 \EE^{\QQ^{(0)}}[\langle N\rangle_{T_H}] < \infty.
\]
Thus $\QQ^{(0)}\in\mathcal M_f$. By \cref{eq:bsde},
\[
d\Psi_t = \big(\tfrac12\theta_t^2 - \tfrac12|\boldsymbol\Lambda^0_t|^2  +  \theta_t\,\mathbf V_t\cdot\mathbf r\big)\,dt  +  \mathbf V_t\cdot d\mathbf B_t.
\]
Since $\mathbf V\cdot d\mathbf B = (\mathbf V\cdot\mathbf r)\,dW^X  +  \boldsymbol\Lambda^0\cdot d\mathbf B$ and $|-\theta\mathbf r  +  \boldsymbol\Lambda^0|^2 = \theta^2  +  |\boldsymbol\Lambda^0|^2$, the logarithm of $\mathcal E(N)_{T_H}$ is
\[
\int_0^{T_H}(-\theta\mathbf r  +  \boldsymbol\Lambda^0)\cdot d\mathbf B - \tfrac12\int_0^{T_H}\big(\theta^2  +  |\boldsymbol\Lambda^0|^2\big)\,dt = -\Psi_0 - \int_0^{T_H}(\theta  +  \mathbf V\cdot\mathbf r)(dW^X  +  \theta\,dt),
\]
using $\Psi_{T_H} = 0$. As $\theta = \bar\mu/\varsigma$, the last integral is $\chi\mathcal G_{T_H}(\pi^\ast)$, which proves the density formula.

Let $\QQ\in\mathcal M_f$. Under $\QQ$, $W^X  +  \int\theta\,dt$ is a Brownian motion by \cref{prop:emm}(ii), so
\[
\chi\mathcal G(\pi^\ast) = \int(\theta  +  \mathbf V\cdot\mathbf r)\,d(W^X  +  \int \theta\,dt)
\]
is a $\QQ$-local martingale. Its quadratic variation is $\langle N'\rangle$, the same under $\QQ$ and $\PP$, where $N' = \int(\theta  +  \mathbf V\cdot\mathbf r)\,dW^X$ is BMO under $\PP$. By the John--Nirenberg inequality, \cite{Kazamaki1994}, there is $\varepsilon > 0$ with $\EE^{\PP}[ \exp( \varepsilon \langle N' \rangle_{T_H})] < \infty$. Young's inequality $xy \le  x\log x - x  +  e^y$, applied with $x = d\QQ/d\PP$ and $y = \varepsilon\langle N'\rangle_{T_H}$, gives
\[
\varepsilon\EE^{\QQ}[\langle N'\rangle_{T_H}] \le  H(\QQ|\PP) - 1  +  \EE^{\PP}[e^{\varepsilon\langle N'\rangle_{T_H}}] < \infty.
\]
By the Burkholder--Davis--Gundy inequality under $\QQ$, $\sup_{t \le  T_H}|\mathcal G_t(\pi^\ast)|$ is $\QQ$-integrable, so $\mathcal G(\pi^\ast)$ is a uniformly integrable $\QQ$-martingale. Hence $\pi^\ast\in\mathcal A$, and taking $\QQ = \QQ^{(0)}$ in the logarithm of the density gives $H(\QQ^{(0)}|\PP) = -\Psi_0$.

(c) Let $\pi\in\mathcal A$. Both $\mathcal G(\pi)$ and $\mathcal G(\pi^\ast)$ are
$\QQ^{(0)}$-martingales with zero expectation, so by (b) and Jensen's inequality
\[
\EE^{\PP}\big[e^{-\chi\mathcal G_{T_H}(\pi)}\big] = e^{\Psi_0}\EE^{\QQ^{(0)}}\big[e^{-\chi(\mathcal G_{T_H}(\pi) - \mathcal G_{T_H}(\pi^\ast))}\big]\ge e^{\Psi_0},
\]
with equality at $\pi = \pi^\ast$.

(d) For $\QQ\in\mathcal M_f$, the logarithm of $d\QQ^{(0)}/d\PP$ is $\QQ$-integrable with $\QQ$-expectation $-\Psi_0$, by (b), so $H(\QQ|\PP) = H(\QQ|\QQ^{(0)}) - \Psi_0$.

(e) The marginal utility at the optimum is proportional to $e^{-\chi\mathcal G_{T_H}(\pi^\ast)}$, hence to $d\QQ^{(0)}/d\PP$. The price of a forward payment $e^{\int_T^{T_H}r_s\,ds}H$ is therefore its $\QQ^{(0)}$-expectation, and conversion to current units gives the stated discounted expectation.
\end{proof}

\section{Proof of \cref{prop:noleverage}} \label{app:theorem7}

\begin{proof}
With $c_j\equiv0$ and $\boldsymbol\Lambda=0$, subtracting the equations for $u_i$ and $\zeta_i$ in \cref{eq:lift_S} gives the closed system
\[
dY_t=\tilde m(Y_t,h_t)\,dt, \qquad dh_{i,t}=-\big(\gamma_ih_{i,t} + a_i\tilde m(Y_t,h_t) + q_i\rho_i\tilde\theta(t,Y_t,h_t)\big)\,dt + q_i\,dB^{\QQ_0}_{i,t},
\]
with $\tilde m(y,h)=-U'(y) + \sum_ih_i$, as in \cref{prop:reduced}. Let $\tilde L$ be its generator. For $\omega=\tilde\omega(t,y,\zeta-u)$ we have $\partial_{\zeta_i}\omega=\partial_{h_i}\tilde\omega=-\partial_{u_i}\omega$, the coordinates $u_i$ carry no diffusion, and $\omega$ does not depend on $x$ or $g$. Hence $\mathcal L^{\QQ_0}\omega=\tilde L\tilde\omega$, $e(\omega)=0$ and $v_i(\omega)=q_i\partial_{h_i}\tilde\omega$. Under either alternative on the correlations, \cref{eq:residual_variance} gives $\mathcal R(\omega)=\kappa_\rho\sum_iq_i^2(\partial_{h_i}\tilde\omega)^2$. For $\tilde\psi=e^{\kappa_\rho\tilde\omega}$ it follows that
\[
\partial_t\tilde\psi + \tilde L\tilde\psi-\tfrac{\kappa_\rho}{2}\tilde\theta^2\tilde\psi=\kappa_\rho\tilde\psi\big(\partial_t\tilde\omega + \tilde L\tilde\omega + \tfrac12\mathcal R(\omega)-\tfrac12\tilde\theta^2\big).
\]
Thus \cref{eq:ptoq_omega} is equivalent to the linear equation $\partial_t\tilde\psi + \tilde L\tilde\psi-\tfrac{\kappa_\rho}{2}\tilde\theta^2\tilde\psi=0$ with $\tilde\psi(T_H,\cdot)=1$. The expectation in \cref{eq:distortion} is
\[
\tilde\psi(t,y,h) = \EE \big[\exp \big(-\tfrac{\kappa_\rho}{2} \int_t^{T_H} \tilde\theta(s,Y_s,h_s)^2\,ds \big) \big]
\]
for the solution started from $(y,h)$ at time $t$, and $e^{-\kappa_\rho \bar\theta^2(T_H-t)/2}  \le  \tilde \psi  \le  1$, which gives the bounds on $\omega$.

It remains to show that $\tilde\psi$ is a classical solution with bounded first derivatives. The noise of $(Y,h)$ is additive. For each path of $\mathbf B^{\QQ_0}$, the processes $Y$ and $h_i-q_i(B^{\QQ_0}_i-B^{\QQ_0}_{i,t})$ therefore solve an ordinary differential equation with coefficients continuous in time and of class $C^2$ in the state, and the solution does not explode by \cref{lem:martingale}. The solution is thus twice continuously differentiable in its initial condition, and its first derivatives $(\delta Y,\delta h)$ in a given direction solve
\[
\dot{\delta Y}=-U''(Y)\,\delta Y + \sum_i\delta h_i, \qquad \dot{\delta h}_i=-\gamma_i\delta h_i-a_i\dot{\delta Y}-q_i\rho_i\nabla\tilde\theta\cdot(\delta Y,\delta h).
\]
In the variables $\delta w_i=\delta h_i + a_i\delta Y$, a direct computation gives
\begin{equation} \label{eq:variation_energy}
\frac{d}{dt}\Big(\tfrac12\delta Y^2 + \sum_i\frac{\delta w_i^2}{2\gamma_ia_i}\Big)=-U''(Y)\,\delta Y^2-\sum_i\Big(\sqrt{a_i}\,\delta Y-\frac{\delta w_i}{\sqrt{a_i}}\Big)^2-\sum_i\frac{q_i\rho_i}{\gamma_ia_i}\,\delta w_i\,\nabla\tilde\theta\cdot(\delta Y,\delta h).
\end{equation}
Since $U''\ge-K$ and $\nabla\tilde\theta$ is bounded, the right side is at most a constant times the energy, however large $U''(Y)$ becomes. Gronwall's inequality therefore bounds the first derivatives of the solution on $[t,T_H]$ by a deterministic constant times the size of the initial perturbation. Differentiating under the expectation shows that the first derivatives of $\tilde\psi$ are bounded. The second derivatives of the solution solve the same linear system with the additional terms $-U'''(Y)\,\delta^{(1)}Y\,\delta^{(2)}Y$ in the equation for $Y$ and $-q_i\rho_i\nabla^2\tilde\theta\big(\delta^{(1)},\delta^{(2)}\big)$ in the equations for $h_i$, where $\delta^{(1)}$ and $\delta^{(2)}$ are first derivatives in two directions.

To control them, let
\[
G_0(y,u,\zeta) = 1 + U(y) - U_\ast + \sum_i u_i^2/(2a_i) + \sum_i \zeta_i^2.
\]
The calculation in the proof of \cref{thm:dichotomy}(iv), with the bounded drift $-q_i\rho_i\tilde\theta$ added to the noise modes, gives $\mathcal L^{\QQ_0}G_0\le CG_0$. Only the $\zeta_i$ carry diffusion, with constant loadings, so
\[
\mathcal L^{\QQ_0}(G_0^2) = 2 G_0 \mathcal L^{\QQ_0}G_0 + 4\sum_i q_i^2 \zeta_i^2 \le CG_0^2.
\]
Stopping on balls and Gronwall's inequality bound $\EE[G_0(Y_s,u_s,\zeta_s)^2]$ uniformly for $s\in[t,T_H]$ and for initial points in compact sets, where a reduced initial point $(y,h)$ is lifted to $u=0$ and $\zeta=h$. By variation of constants and \cref{eq:variation_energy}, the second derivatives of the solution are bounded by $C\int_t^{T_H}\big(1 + |U'''(Y_s)|\big)\,ds$, and
\[
|U'''|\le K(1 + U-U_\ast) \le K G_0.
\]
They are therefore bounded in $L^2$, locally uniformly in the initial point. This uniform integrability allows two differentiations under the expectation and the passage to the limit in the second derivatives, which are therefore continuous and bounded on compact sets uniformly in $t$. The Markov property and It\^o's formula then show that $\tilde\psi$ is continuously differentiable in $t$ and solves the linear equation, and $\tilde\psi(t,\cdot) \to 1$ uniformly as $t \to  T_H$. Hence $\tilde \omega = \kappa_\rho^{-1} \log \tilde \psi$ has the stated regularity and solves \cref{eq:ptoq_omega}. Its exposures $v_i( \omega) = q_i \partial_{h_i} \tilde \psi/(\kappa_\rho \tilde\psi)$ are bounded because $\tilde\psi$ is bounded below. The premium \cref{eq:ptoq_mpr} is linear in the $v_i(\omega)$, so it is bounded and locally Lipschitz in $\mathbf z$ uniformly in $t$.
\end{proof}

\section{Proof of \cref{lem:continuity}} \label{app:lemma}

\begin{proof}
Realize all solutions of \cref{eq:lift_S} on one probability space with fixed Brownian motions $\mathbf B^{\QQ}$. Let $(L_n,\vartheta_n) \to (L,\vartheta)$, and let $\Zs^n$ and $\Zs$ be the corresponding solutions. The coefficients converge uniformly on compact sets, and on each ball they are Lipschitz with constants independent of $n$. By \cref{lem:martingale} and (B1), the stopped Lyapunov estimates are uniform in $n$. In particular, if $\tau_R^n=\inf\{t:|\Zs_t^n|\ge R\}$, then $\sup_n\EE[G_{\vartheta_n}(\Zs^n_{T^\ast\wedge\tau_R^n})] \le  C$.

The uniform coercivity in (B1) therefore gives
\[
\sup_n \QQ(\tau_R^n \le  T^\ast) \le  C/(c_\ast R^2-C_\ast)  \to  0.
\]
The same estimate holds for the limiting system. Let $\sigma_R = \inf\{t:|\Zs^n_t|\vee|\Zs_t|\ge R\}$, let $K_R$ be a common Lipschitz constant of the coefficients on the ball of radius $R$, and let $\delta_n(R)$ be the uniform distance between the coefficients of the two systems on $[0,T^\ast]$ times that ball. The Burkholder--Davis--Gundy and Cauchy--Schwarz inequalities give, for
$\Xi_n(t) = \EE\sup_{s \le  t\wedge\sigma_R}|\Zs^n_s - \Zs_s|^2$,
\[
\Xi_n(t) \le  C_R\Big(\delta_n(R)^2  +  \int_0^t\Xi_n(s)\,ds\Big),
\]
so $\Xi_n(T^\ast) \le  C_R\delta_n(R)^2e^{C_RT^\ast}$ by Gronwall's inequality.

Hence
\[
\QQ(\sup_{t \le  T^\ast}|\Zs^n_t - \Zs_t| > \varepsilon) \le \QQ(\sigma_R \le  T^\ast)  +  \varepsilon^{-2}C_R\delta_n(R)^2e^{C_RT^\ast},
\]
and choosing first $R$ and then $n$ large shows that $\sup_{t \le  T^\ast}|\Zs^n_t - \Zs_t| \to  0$ in probability.

\myparagraph{Calls.}
The discounted total-return values $e^{-\int_0^{T_k}(r - d)\,du}S^n_{T_k}$ are nonnegative, converge in probability, and have the same expectation $S_0$ for every $n$, by \cref{lem:martingale}. By Scheff\'e's lemma they converge in $L^1$, and the call payoffs, which are dominated by the price, converge in $L^1$.

\myparagraph{VIX instruments.}
The argument above, started at time $T_m$ from points $\mathbf z_n \to \mathbf z$, shows that $\int_{T_m}^{T_m  +  \Delta}(\varsigma^n_u)^2\,du$ converges in probability. Since $\varsigma^4 \le  CG_{\vartheta}$, these integrals are bounded in $L^2$ uniformly on compact sets of initial points, hence uniformly integrable. Thus $(\mathbf z,L,\vartheta)\mapsto\mathcal V_{T_m}(\mathbf z;L,\vartheta)$ is sequentially continuous. Along any subsequence there is a further subsequence with $\Zs^n_{T_m} \to \Zs_{T_m}$ almost surely, so $\mathcal V_{T_m}(\Zs^n_{T_m};L_n,\vartheta_n) \to \mathcal V_{T_m}(\Zs_{T_m};L,\vartheta)$ in probability. By Jensen's inequality,
\[
\EE[\mathcal V_{T_m}(\Zs^n_{T_m})^2]  \le  \Delta\, \EE \int_{T_m}^{T_m  +  \Delta}(\varsigma^n_u)^4\,du \le  C, \qquad H_m(v)| \le  C(1  +  v)
\]
with $\mathrm{VIX}^2 = \mathcal V/\Delta$, so the payoffs are uniformly integrable and their expectations converge.
\end{proof}

\end{document}